\documentclass[12pt]{article}
\usepackage{myart}

\renewcommand{\theequation}{\arabic{section}.\arabic{equation}}
\normalbaselines
\input tcilatex.tex

\begin{document}

\author{Yuri A. Rylov}
\title{Is the Dirac particle completely relativistic?}
\date{Institute for Problems in Mechanics, Russian Academy of Sciences \\
101-1 ,Vernadskii Ave., Moscow, 119526, Russia \\
email: rylov@ipmnet.ru\\
Web site: {$http://rsfq1.physics.sunysb.edu/\symbol{126}rylov/yrylov.htm$}\\
or mirror Web site: {$http://195.208.200.111/\symbol{126}rylov/yrylov.htm$}}
\maketitle

\begin{abstract}
The Dirac particle, i.e. the dynamic system $\mathcal{S}_{\mathrm{D}}$,
described by the free Dirac equation is investigated. Although the Dirac
equation is written usually in the relativistically covariant form, the
dynamic system $\mathcal{S}_{\mathrm{D}}$ is not completely relativistic,
because its description contains such absolute objects as $\gamma $-matrices 
$\gamma ^{k}$, forming a matrix vector. By means of the proper change of
variables the $\gamma $-matrices are eliminated, but instead of them the
constant timlike vector $f^{k}$ appears. The vector $f^{k}$ describes an
absolute splitting of the space-time into space and time, which is
characteristic for the nonrelativistic description. To investigate a degree
of the violation of the $\mathcal{S}_{\mathrm{D}}$ relativistic description,
we consider the classical Dirac particle $\mathcal{S}_{\mathrm{Dcl}}$,
obtained from $\mathcal{S}_{\mathrm{D}}$ by means of the relativistic
dynamic disquantization. The classical dynamic system $\mathcal{S}_{\mathrm{%
Dcl}}$ appears to be composite, because it has ten degrees of freedom. Six
translational degrees of freedom are described relativistically (without a
reference to $f^{k}$), whereas four internal degrees of freedom are
described nonrelativistically, because their description refers to $f^{k}$.
Coupling the absolute vector $f^{k}$ with the energy-momentum vector of $%
\mathcal{S}_{\mathrm{Dcl}}$, the classical Dirac particle $\mathcal{S}_{%
\mathrm{Dcl}}$ is modified minimally. The vector $f^{k}$ ceases to be
absolute, and the modified classical Dirac particle $\mathcal{S}_{\mathrm{%
mDcl}}$ becomes to be completely relativistic. The dynamic equations for $%
\mathcal{S}_{\mathrm{mDcl}}$ are solved. Solutions for $\mathcal{S}_{\mathrm{%
Dcl}}$ and $\mathcal{S}_{\mathrm{mDcl}}$ are compared.
\end{abstract}


\newpage

\section{Introduction}

The question-mark in the title of the paper seems to be inadequate, because
the problem of the relativistic invariance of the Dirac equation had been
solved many years ago. The Dirac equation was invented by Dirac as a
relativistic equation, and its relativistic invariance was proved by Dirac 
\cite{D58}. The Dirac equation was investigated by many authors \cite{FW50}
- \cite{RV93}, but this investigation was produced always in the framework
of the quantum principles. The proof of the relativistic invariance of the
Dirac equation can be found in any textbook on quantum mechanics. Why does
this question arise?

The fact is that, in general, the relativistic covariance of dynamic
equations is not sufficient for compatibility with the relativity
principles. The Dirac equation contains such specific quantities as $\gamma $%
-matrices. If we make a change of variables, eliminating $\gamma $-matrices,
we obtain the system of dynamic equations, which is not relativistically
covariant. Such a transition to eight hydrodynamic variables: 4-vector $%
j^{k}=\bar{\psi}\gamma ^{k}\psi $, $k=0,1,2,3$ , 4-pseudovector $S^{k}=i\bar{%
\psi}\gamma _{5}\gamma ^{k}\psi $, $k=0,1,2,3$, scalar $\varphi $ and
pseudoscalar $\kappa $ instead of eight real dependent variables $\psi $ is
necessary, if we want to obtain classical analog $\mathcal{S}_{\mathrm{Dcl}}$
of the Dirac particle $\mathcal{S}_{\mathrm{D}}$ \cite{R2004}.

The Dirac particle $\mathcal{S}_{\mathrm{D}}$ is the dynamic system which is
described by the action 
\begin{equation}
\mathcal{S}_{\mathrm{D}}:\qquad \mathcal{A}_{\mathrm{D}}[\bar{\psi},\psi
]=\int (-m\bar{\psi}\psi +\frac{i}{2}\hbar \bar{\psi}\gamma ^{l}\partial
_{l}\psi -\frac{i}{2}\hbar \partial _{l}\bar{\psi}\gamma ^{l}\psi )d^{4}x
\label{b1.1}
\end{equation}
Here $\psi $ is four-component complex wave function, $\psi ^{\ast }$ is the
Hermitian conjugate wave function, and $\bar{\psi}=\psi ^{\ast }\gamma ^{0}$
is conjugate one. $\gamma ^{i}$, $i=0,1,2,3$ are $4\times 4$ complex
constant matrices, satisfying the relation 
\begin{equation}
\gamma ^{l}\gamma ^{k}+\gamma ^{k}\gamma ^{l}=2g^{kl}I,\qquad k,l=0,1,2,3.
\label{b1.2}
\end{equation}
where $I$ is the unit $4\times 4$ matrix, and $g^{kl}=$diag$\left(
c^{-2},-1,-1,-1\right) $ is the metric tensor. Considering dynamic system $%
\mathcal{S}_{\mathrm{D}}$, we choose for simplicity such units, where the
speed of the light $c=1$. The action (\ref{b1.1}) generates dynamic equation 
\begin{equation}
i\hbar \gamma ^{l}\partial _{l}\psi -m\psi =0  \label{f1.2}
\end{equation}
and expressions for physical quantities: the 4-current $j^{k}$ of particles
and the energy-momentum tensor $T_{l}^{k}$%
\begin{equation}
j^{k}=\bar{\psi}\gamma ^{k}\psi ,\qquad T_{l}^{k}=\frac{i}{2}\left( \bar{\psi%
}\gamma ^{k}\partial _{l}\psi -\partial _{l}\bar{\psi}\cdot \gamma ^{k}\psi
\right)  \label{f1.3}
\end{equation}
The dynamic equation (\ref{f1.2}) is known as the Dirac equation.

We stress that the current $j^{k}$, as well as the energy-momentum tensor $%
T_{l}^{k}$ are attributes of the Dirac particle $\mathcal{S}_{\mathrm{D}}$.
In particular, it means that, changing expression (\ref{f1.3}) for the
current $j^{k}$, we change the dynamic system $\mathcal{S}_{\mathrm{D}}$,
even if the dynamic equation (\ref{f1.2}) is not changed.

The classical Dirac particle $\mathcal{S}_{\mathrm{Dcl}}$ (classical analog
of $\mathcal{S}_{\mathrm{D}}$) is a discrete dynamic system, i.e. the
dynamic system, having finite number (ten) of the freedom degrees. Action
and dynamic equations for $\mathcal{S}_{\mathrm{Dcl}}$ are obtained as a
result of dynamic disquantization of the dynamic system $\mathcal{S}_{%
\mathrm{D}}$ \cite{R2004}. By definition the procedure of dynamic
disquantization means the change 
\begin{equation}
\partial ^{l}\rightarrow \partial _{||}^{l}=\frac{j^{l}j^{k}}{j^{s}j_{s}}%
\partial _{k},\qquad l=0,1,2,3,\qquad \partial ^{k}\equiv g^{kl}\partial
_{l}\equiv g^{kl}\frac{\partial }{\partial x_{l}},\qquad j_{l}\equiv
g_{lk}j^{k}  \label{b1.8}
\end{equation}
in the action (\ref{b1.1}) and, hence, in the dynamic equation (\ref{f1.2}).

After such a change the system of partial differential equations (\ref{f1.2}%
) becomes to be equivalent to a system of ordinary differential equations,
because all dynamic equations contain derivatives only in the direction of
the vector $j^{k}$. The field of vector $j^{k}$ in the space-time determines
a set of world lines $\mathcal{L}$ tangent to $j^{k}$. After dynamic
disquantization (\ref{b1.8}) the dynamic equations contain only derivatives
along $\mathcal{L}$. It means that giving initial values of dependent
dynamic variables at some point of the world line $\mathcal{L}_{1}$, one can
determine by means of dynamic equations the values of the dependent dynamic
variables at all points of $\mathcal{L}_{1}$ independently of other world
lines $\mathcal{L}$. It means that after dynamic disquantization the dynamic
system $\mathcal{S}_{\mathrm{D}}$ turns into the dynamic system $\mathcal{S}%
_{\mathrm{Dqu}}$, describing the statistical ensemble $\mathcal{E}\left[ 
\mathcal{S}_{\mathrm{Dcl}}\right] $ of classical dynamic system $\mathcal{S}%
_{\mathrm{Dcl}}$.

Dynamic disquantization has the following properties:

\begin{enumerate}
\item  Dynamic disquantization is determined completely by the dynamic
system $\mathcal{S}_{\mathrm{D}}$.

\item  Dynamic disquantization leads to the unique resulting dynamic system $%
\mathcal{S}_{\mathrm{Dcl}}$.

\item  Dynamic disquantization is a relativistically covariant procedure.
\end{enumerate}

It follows from the two last properties, that if the dynamic system $%
\mathcal{S}_{\mathrm{Dcl}}$ is nonrelativistic, the dynamic system $\mathcal{%
S}_{\mathrm{D}}$ cannot be relativistic.

In general, using the first equation (\ref{f1.3}), we can apply the
procedure (\ref{b1.8}) directly to (\ref{f1.2}). But the obtained system of
ordinary differential equations 
\[
i\hbar \left( \bar{\psi}\gamma _{l}\psi \right) \left( \bar{\psi}\gamma
^{k}\psi \right) \gamma ^{l}\partial _{k}\psi -m\left( \bar{\psi}\gamma
_{k}\psi \right) \left( \bar{\psi}\gamma ^{k}\psi \right) \psi =0 
\]
is too complicated for investigation, because it is nonlinear. Description
in terms of the wave function is effective only, if the dynamic equations
are linear in terms of the wave function. In the case of nonlinear dynamic
equations it is more effective to introduce dynamic variables, whose
physical meaning is clear.

Investigation of the dynamic system $\mathcal{S}_{\mathrm{Dcl}}$ instead of $%
\mathcal{S}_{\mathrm{D}}$ is useful, because $\mathcal{S}_{\mathrm{Dcl}}$
has finite number of the freedom degrees. It may appear that some of degrees
of freedom are described relativistically, but another ones are described
nonrelativistically.

\section{Transformation of variables}

The state of dynamic system $\mathcal{S}_{\mathrm{D}}$ is described by eight
real dependent variables (eight real components of four-component complex
wave function $\psi $). Transforming the action (\ref{b1.1}), we use the
mathematical technique \cite{S30,S51}, where the wave function $\psi $ is
considered to be a function of hypercomplex numbers $\gamma $ and
coordinates $x$. In this case the dynamic quantities are obtained by means
of a convolution of expressions $\psi ^{\ast }O\psi $ with zero divisors.
This technique allows one to work without fixing the $\gamma $-matrices
representation.

Using designations 
\begin{equation}
\gamma _{5}=\gamma ^{0123}\equiv \gamma ^{0}\gamma ^{1}\gamma ^{2}\gamma
^{3},  \label{f1.9}
\end{equation}
\begin{equation}
\mathbf{\sigma }=\{\sigma _{1},\sigma _{2},\sigma _{3},\}=\{-i\gamma
^{2}\gamma ^{3},-i\gamma ^{3}\gamma ^{1},-i\gamma ^{1}\gamma ^{2}\}
\label{f1.10}
\end{equation}
we make the change of variables 
\begin{equation}
\psi =Ae^{i\varphi +{\frac{1}{2}}\gamma _{5}\kappa }\exp \left( -\frac{i}{2}%
\gamma _{5}\mathbf{\sigma \eta }\right) \exp \left( {\frac{i\pi }{2}}\mathbf{%
\sigma n}\right) \Pi  \label{f1.11}
\end{equation}
\[
\bar{\psi}=\psi ^{\ast }\gamma ^{0} 
\]
\begin{equation}
\psi ^{\ast }=A\Pi \exp \left( -{\frac{i\pi }{2}}\mathbf{\sigma n}\right)
\exp \left( -\frac{i}{2}\gamma _{5}\mathbf{\sigma \eta }\right) e^{-i\varphi
-{\frac{1}{2}}\gamma _{5}\kappa }  \label{f1.12}
\end{equation}
where (*) means the Hermitian conjugation, and 
\begin{equation}
\Pi ={\frac{1}{4}}(1+\gamma ^{0})(1+\mathbf{z\sigma }),\qquad \mathbf{z}%
=\{z^{\alpha }\}=\text{const},\qquad \alpha =1,2,3;\qquad \mathbf{z}^{2}=1
\label{f1.13}
\end{equation}
is a zero divisor. The quantities $A$, $\kappa $, $\varphi $, $\mathbf{\eta }%
=\{\eta ^{\alpha }\}$, $\mathbf{n}=\{n^{\alpha }\}$, $\alpha =1,2,3,\;$ $%
\mathbf{n}^{2}=1$ are eight real parameters, determining the wave function $%
\psi $. These parameters may be considered as new dependent variables,
describing the state of dynamic system $\mathcal{S}_{\mathrm{D}}$. The
quantity $\varphi $ is a scalar, and $\kappa $ is a pseudoscalar. Six
remaining variables $A,$ $\mathbf{\eta }=\{\eta ^{\alpha }\}$, $\mathbf{n}%
=\{n^{\alpha }\}$, $\alpha =1,2,3,\;$ $\mathbf{n}^{2}=1$ can be expressed
through the flux 4-vector $j^{l}=\bar{\psi}\gamma ^{l}\psi $ and the spin
4-pseudovector 
\begin{equation}
S^{l}=i\bar{\psi}\gamma _{5}\gamma ^{l}\psi ,\qquad l=0,1,2,3  \label{f1.13a}
\end{equation}
Because of two identities 
\begin{equation}
S^{l}S_{l}\equiv -j^{l}j_{l},\qquad j^{l}S_{l}\equiv 0.  \label{f1.14}
\end{equation}
there are only six independent components among eight components of
quantities $j^{l}$, and $S^{l}$.

After transformation we obtain (see details in \cite{R2004})

\begin{equation}
\mathcal{S}_{\mathrm{D}}:\qquad \mathcal{A}_{\mathrm{D}}[j,\varphi ,\kappa ,%
\mathbf{\xi }]=\int \mathcal{L}d^{4}x,\qquad \mathcal{L}=\mathcal{L}_{%
\mathrm{cl}}+\mathcal{L}_{\mathrm{q1}}+\mathcal{L}_{\mathrm{q2}}
\label{c4.15}
\end{equation}
\begin{equation}
\mathcal{L}_{\mathrm{cl}}=-m\rho -\hbar j^{i}\partial _{i}\varphi -\frac{%
\hbar j^{l}}{2\left( 1+\mathbf{\xi z}\right) }\varepsilon _{\alpha \beta
\gamma }\xi ^{\alpha }\partial _{l}\xi ^{\beta }z^{\gamma },\qquad \rho
\equiv \sqrt{j^{l}j_{l}}  \label{c4.16}
\end{equation}
\begin{equation}
\mathcal{L}_{\mathrm{q1}}=2m\rho \sin ^{2}(\frac{\kappa }{2})-{\frac{\hbar }{%
2}}S^{l}\partial _{l}\kappa ,  \label{c4.17}
\end{equation}
\begin{equation}
\mathcal{L}_{\mathrm{q2}}=\frac{\hbar (\rho +j_{0})}{2}\varepsilon _{\alpha
\beta \gamma }\partial ^{\alpha }\frac{j^{\beta }}{(j^{0}+\rho )}\xi
^{\gamma }-\frac{\hbar }{2(\rho +j_{0})}\varepsilon _{\alpha \beta \gamma
}\left( \partial ^{0}j^{\beta }\right) j^{\alpha }\xi ^{\gamma }
\label{c4.18}
\end{equation}
Here and in what follows, a summation is produced over repeated Greek
indices (1-3) and over repeated Latin indices (0-3). Lagrangian is a
function of 4-vector $j^{l}$, scalar $\varphi $, pseudoscalar $\kappa $, and
unit 3-pseudovector $\mathbf{\xi }$, which is connected with the spin
4-pseudovector $S^{l}$ by means of the relations 
\begin{equation}
\xi ^{\alpha }=\rho ^{-1}\left( S^{\alpha }-\frac{j^{\alpha }S^{0}}{%
(j^{0}+\rho )}\right) ,\qquad \alpha =1,2,3;\qquad \rho \equiv \sqrt{%
j^{l}j_{l}}  \label{f1.15}
\end{equation}
\begin{equation}
S^{0}=\mathbf{j\xi },\qquad S^{\alpha }=\rho \xi ^{\alpha }+\frac{(\mathbf{%
j\xi })j^{\alpha }}{\rho +j^{0}},\qquad \alpha =1,2,3  \label{f1.16}
\end{equation}
The unit 3-pseudovector $\mathbf{\xi }=\left\{ \xi _{1},\xi _{2},\xi
_{3}\right\} $ is connected with the 3-vector $\mathbf{n}=\left\{
n_{1},n_{2},n_{3},\right\} $ in (\ref{f1.11}), (\ref{f1.12})\textbf{\ }by
means of the relation 
\begin{equation}
\mathbf{\xi }=2\mathbf{n}(\mathbf{nz})-\mathbf{z}  \label{a3.21}
\end{equation}

As one can see from (\ref{c4.16}) -- (\ref{c4.18}), only two first terms in (%
\ref{c4.16}) and two term in (\ref{c4.17}) are written in the
relativistically covariant form. They are invariants. Other terms are
written in the non-covariant form. The non-covariant term in (\ref{c4.16})
contains three-dimensional Levi-Chivita pseudotensor $\varepsilon _{\alpha
\beta \gamma }$. It can be considered as spatial components of the
4-dimensional Levi-Chivita pseudotensor $\varepsilon _{iklm}\;\;(\varepsilon
_{0123}=1)$, convoluted with the constant timelike unit vector $%
f^{l}=\{1,0,0,0\}$. Then only spatial components of pseudotensor $%
\varepsilon _{iklm}f^{m}$ do not vanish 
\begin{equation}
\varepsilon _{\alpha \beta \gamma }=-\varepsilon _{\alpha \beta \gamma
m}f^{m},\qquad \alpha ,\beta ,\gamma =1,2,3  \label{b5.12}
\end{equation}
and one may substitute relation (\ref{b5.12}) in expression (\ref{c4.16}).

To write the expression (\ref{c4.18}) in the relativistically covariant
form, we also introduce constant 4-vector $f^{l}$ 
\begin{equation}
f^{i}=\{1,0,0,0\}  \label{d2.1}
\end{equation}
and take into account that the three-dimensional Levi-Chivita pseudotensor $%
\varepsilon _{\alpha \beta \gamma }$ may be considered as the component $%
\varepsilon _{0\alpha \beta \gamma }$ of the 4-dimensional Levi-Chivita
pseudotensor $\varepsilon _{iklm}$. It easy to verify that the expression (%
\ref{c4.18}) may be written in the form 
\begin{equation}
\mathcal{L}_{\mathrm{q2}}=\frac{\hbar }{2}\varepsilon _{ikl\gamma }\left(
j^{i}+f^{i}\rho \right) \partial ^{k}\frac{j^{l}+f^{l}\rho }{%
(j^{s}f_{s}+\rho )}\xi ^{\gamma }  \label{d2.2}
\end{equation}
where summation is produced over Latin indices $\left( 0-3\right) $ and over
Greek indices $\left( 1-3\right) $.

Let us write relations (\ref{f1.15}), (\ref{f1.16}) in the covariant form.
We shall consider 3-pseudovector $\mathbf{\xi }$ $=\left\{ \xi ^{1},\xi
^{2},\xi ^{3}\right\} $ as spatial components of 4-pseudovector $\xi
^{k}=\left\{ \xi ^{0},\mathbf{\xi }\right\} =\left\{ \xi ^{0},\xi ^{1},\xi
^{2},\xi ^{3}\right\} $. Then relations (\ref{f1.15}), (\ref{f1.16}) take
the form 
\begin{equation}
\xi ^{k}=\rho ^{-1}\left( S^{k}-\frac{j^{k}S^{l}f_{l}}{(j^{s}f_{s}+\rho )}%
\right) ,\qquad k=0,1,2,3;  \label{d2.3}
\end{equation}
\[
\xi ^{k}\xi _{k}=\left( \frac{S^{l}f_{l}}{(j^{s}f_{s}+\rho )}\right) ^{2}-1 
\]
\begin{equation}
S^{k}=\xi ^{k}\rho +j^{k}\left( \xi ^{s}f_{s}\right) ,\qquad k=0,1,2,3
\label{d2.4}
\end{equation}

We introduce the four-component quantity 
\begin{equation}
\nu ^{k}=\xi ^{k}-f^{k}\left( \xi ^{s}f_{s}\right) ,\qquad k=0,1,2,3;\qquad
\nu ^{l}\nu _{l}=-1  \label{d2.5}
\end{equation}
In the coordinate system, where the expression (\ref{d2.2}) is written and
where the vector $f^{k}$ has the form (\ref{d2.1}), we obtain $\nu ^{0}=0$, $%
\nu ^{\alpha }=\xi ^{\alpha }$, $\alpha =1,2,3$. The relation (\ref{d2.2})
can be written in the covariant form 
\begin{equation}
\mathcal{L}_{\mathrm{q2}}=\hbar \rho \varepsilon _{iklm}q^{i}\left( \partial
^{k}q^{l}\right) \nu ^{m}  \label{d2.6}
\end{equation}
where $q^{k}$ is the unit 4-vector 
\begin{equation}
q^{k}=\frac{j^{k}+f^{k}\rho }{\sqrt{2\rho \left( j^{s}f_{s}+\rho \right) }}%
,\qquad q^{k}q_{k}=1  \label{d2.7}
\end{equation}

Analogously, one can write the expression (\ref{c4.16}) in the covariant
form. Introducing the unit spacelike constant 4-pseudovector 
\begin{equation}
z^{k}=\left\{ 0,\mathbf{z}\right\} =\left( 0,z^{1},z^{2},z^{3}\right)
,\qquad z^{k}z_{k}=-1,\qquad \mathbf{\xi z}=-\xi _{l}z^{l}  \label{d2.8}
\end{equation}
where the 3-pseudovector $\mathbf{z}$ is defined by the relation (\ref{f1.13}%
), we write the last term of the expression (\ref{c4.16}) in the form 
\begin{equation}
-\frac{\hbar j^{l}}{2\left( 1+\mathbf{\xi z}\right) }\varepsilon _{\alpha
\beta \gamma }\xi ^{\alpha }\partial _{l}\xi ^{\beta }z^{\gamma }=-\hbar
j^{s}\varepsilon _{iklm}\frac{\xi ^{i}}{\sqrt{2\left( 1-\xi _{s}z^{s}\right) 
}}\partial _{s}\frac{\xi ^{k}}{\sqrt{2\left( 1-\xi _{s}z^{s}\right) }}%
z^{l}f^{m}  \label{b5.12a}
\end{equation}

The factor $\left[ 2\left( 1-\xi ^{s}z_{s}\right) \right] ^{-1/2}$ is
introduced under sign of derivative, because differentiation of it gives $0$
in virtue of the vanishing factor $\varepsilon _{iklm}\xi ^{i}\xi
^{k}z^{l}f^{m}=0$.

Finally, introducing 4-pseudovector 
\begin{equation}
\mu ^{i}\equiv \frac{\nu ^{i}}{\sqrt{-(\nu ^{l}+z^{l})(\nu _{l}+z_{l})}}=%
\frac{\nu ^{i}}{\sqrt{2(1-\nu ^{l}z_{l})}}=\frac{\nu ^{i}}{\sqrt{2(1+\mathbf{%
\xi z})}}.  \label{b5.12b}
\end{equation}
we can rewrite the last term of (\ref{c4.16}) in the form 
\begin{equation}
-\frac{\hbar j^{l}}{2\left( 1+\mathbf{\xi z}\right) }\varepsilon _{\alpha
\beta \gamma }\xi ^{\alpha }\partial _{l}\xi ^{\beta }z^{\gamma }=\hbar
j^{s}\varepsilon _{iklm}\mu ^{i}\partial _{s}\mu ^{k}z^{l}f^{m}
\label{b5.16}
\end{equation}

Now we can write the action (\ref{c4.15}) in the covariant form 
\begin{equation}
\mathcal{S}_{\mathrm{D}}:\qquad \mathcal{A}_{\mathrm{D}}[j,\varphi ,\kappa ,%
\mathbf{\xi }]=\int \mathcal{L}d^{4}x,\qquad \mathcal{L}=\mathcal{L}_{%
\mathrm{cl}}+\mathcal{L}_{\mathrm{q1}}+\mathcal{L}_{\mathrm{q2}}
\label{c4.15a}
\end{equation}
\begin{equation}
\mathcal{L}_{\mathrm{cl}}=-m\rho -\hbar j^{i}\partial _{i}\varphi +\hbar
j^{s}\varepsilon _{iklm}\mu ^{i}\partial _{s}\mu ^{k}z^{l}f^{m},\qquad \rho
\equiv \sqrt{j^{l}j_{l}}  \label{c4.16a}
\end{equation}
\begin{equation}
\mathcal{L}_{\mathrm{q1}}=2m\rho \sin ^{2}(\frac{\kappa }{2})-{\frac{\hbar }{%
2}}S^{l}\partial _{l}\kappa ,  \label{c4.17a}
\end{equation}
\begin{equation}
\mathcal{L}_{\mathrm{q2}}=\hbar \rho \varepsilon _{iklm}q^{i}(\partial
^{k}q^{l})\nu ^{m}  \label{c4.18a}
\end{equation}
where $\varphi $ is a scalar, $\kappa $ is a pseudoscalar, the quantities $%
j^{k}$, $q^{k}$,$f^{k}$ are 4-vectors and the quantities $S^{k}$, $\mu ^{k}$%
, $\nu ^{k}$, $z^{k}$ are 4-pseudovectors. Vector $S^{k}$ is expressed via
variables $j^{k}$ and $\xi ^{k}$ by means of relations (\ref{d2.4}).
4-pseudovector $\xi ^{k}$ contains only two independent components, because
it satisfies the constraints 
\begin{equation}
\xi ^{k}\xi _{k}=\left( \xi ^{s}f_{s}\right) ^{2}-1,\qquad \left( j_{k}\xi
^{k}\right) +\left( \xi ^{s}f_{s}\right) \rho =0  \label{c4.19}
\end{equation}
which ensure fulfilment of relations (\ref{f1.14}).

\section{Relativistic invariance}

It is a common practice to think that if dynamic equations of a dynamic
system can be written in the relativistically covariant form, such a
possibility provides automatically relativistic character of considered
dynamic system, described by these equations. In general, it is valid only
in the case, when dynamic equations do not contain absolute objects, or
these absolute objects has the Lorentz group as a group of their symmetry 
\cite{A67}. The absolute object is one or several quantities, which are the
same for all states of the dynamic system \cite{A67}. A given external
field, or metric tensor (when it is given, but not determined from the
gravitational equations) are examples of absolute objects. In the case of
dynamic system $\mathcal{S}_{\mathrm{D}}$, described by the action (\ref
{b1.1}) the Dirac $\gamma $-matrices are absolute objects.

Anderson \cite{A67} investigated in details the role of absolute objects for
symmetry of dynamic systems. His conclusion is as follows. If a dynamic
system is described by dynamic equations, written in the covariant form, the
symmetry group of the dynamic system is determined by the symmetry group of
these absolute objects. Here we confirm this result in a simple example,
when the dynamic equations of the certainly nonrelativistic dynamic system
are written in a relativistically covariant form.

Let us consider a system of differential equations, consisting of the
Maxwell equations for the electromagnetic tensor $F^{ik}$ in some inertial
coordinate system

\begin{equation}
\partial _{k}F^{ik}\left( x\right) =4\pi J^{i},\qquad \varepsilon
_{iklm}g^{jm}\partial _{j}F^{kl}\left( x\right) =0,\qquad \partial
_{k}\equiv \frac{\partial }{\partial x^{k}}  \label{a6.1}
\end{equation}
and equations 
\begin{equation}
m\frac{d}{d\tau }\left[ \left( l_{k}\dot{q}^{k}\right) ^{-1}\dot{q}^{i}-{%
\frac{1}{2}}g^{ik}l_{k}\left( l_{j}\dot{q}^{j}\right) ^{-2}\dot{q}^{s}g_{sl}%
\dot{q}^{l}\right] =eF^{il}\left( q\right) g_{lk}\dot{q}^{k};\qquad i=0,1,2,3
\label{a6.2}
\end{equation}
\[
\dot{q}^{k}\equiv \frac{dq^{k}}{d\tau } 
\]
where $q^{i}=q^{i}(\tau )$, $i=0,1,2,3$ describe coordinates of a pointlike
charged particle as functions of a parameter $\tau $, the quantity $l_{k}$, $%
k=0,1,2,3$ is a constant timelike unit vector, 
\begin{equation}
g^{ik}l_{i}l_{k}=1;  \label{a6.3}
\end{equation}
and the speed of the light $c=1$.

This system of equations is relativistically covariant with respect to
quantities $q^{i}$, $F^{ik}$, $J^{i}$, $l_{i}$, $g_{ik}$, i.e. it does not
change its form at any Lorentz transformation, which is accompanied by
corresponding transformation of quantities $q^{i}$, $F^{ik}$, $J^{i}$, $%
l_{i} $, $g_{ik}$, where the quantities $q^{i}$, $J^{i}$, $l_{i}$ are
transformed as 4-vectors and the quantities $F^{ik}$, $g_{ik}$ are
transformed as 4-tensors

The reference to the quantities $q^{i}$, $F^{ik}$, $J^{i}$, $l_{i}$, $g_{ik}$
means that all these quantities are considered as formal dependent
variables, when one compares the form of dynamic equations written in two
different coordinate systems. For instance, if the reference to $J^{i}$ is
omitted in the formulation of the relativistic covariance, it means that
components of $J^{i}$ are considered as some functions of the coordinates $x$%
. Let $J^{i}\neq 0$, and $\tilde{J}^{i}$ be the quantity $J^{i}$, written in
other coordinate system. Then, in general, $J^{i}$ and $\tilde{J}^{i}$ are
different functions of the arguments $x$ and $\tilde{x}$ respectively, and
the first equation (\ref{a6.1}) has different form in different coordinate
systems. In other words, the dynamic equations (\ref{a6.1})--(\ref{a6.2})
are not relativistically covariant, in general, with respect to quantities $%
q^{i}$, $F^{ik}$, $l_{i}$, $g_{ik}$, if $J^{i}\neq 0$. Thus, for the
relativistic covariance it is important both the laws of transformation and
how each of quantities is considered as a formal variable, or as some
function of coordinates.

Following Anderson \cite{A67}, we divide the quantities $q^{i}$, $F^{ik}$, $%
J^{i}$, $l_{i}$, $g_{ik}$ into two parts: dynamic objects (variables) $q^{i}$%
, $F^{ik}$ and absolute objects $J^{i}$, $l_{i}$, $g_{ik}$. According to
definition of absolute objects they have the same value for all solutions of
the dynamic equations, whereas dynamic variables are different, in general,
for different solutions. If the dynamic equations are written in the
relativistically covariant form, their symmetry group is determined by the
symmetry group of the absolute objects $J^{i}$, $l_{i}$, $g_{ik}$. Dynamic
equations are compatible with the principles of relativity if the Lorentz
group is the symmetry group of the absolute objects.

Let for simplicity $J^{i}\equiv 0$. The symmetry group of the constant
timelike vector $l_{i}$ is a group of rotations in the $3$-plane orthogonal
to the vector $l_{i}$. The Lorentz group is a symmetry group of the metric
tensor $g_{ik}=$diag $\{1,-1,-1,-1\}$. Thus, the symmetry group of all
absolute objects $l_{i}$, $g_{ik}$, $J^{i}\equiv 0$ is a subgroup of the
Lorentz group (rotations in the $3$-plane orthogonal to $l_{i}$). As far as
the symmetry group is a subgroup of the Lorentz group and does not coincide
with it, the system of equations (\ref{a6.1})--(\ref{a6.2}) is
nonrelativistic (incompatible with the relativity principles).

Of course, the compatibility with the relativity principles does not depend
on the fact with respect to which quantities the relativistic covariance is
considered. For instance, let us consider a covariance of equations (\ref
{a6.1}), (\ref{a6.2}) with respect to the quantities $q^{i}$, $F^{ik}$, $%
J^{i}\equiv 0$. It means that now the quantities $l_{i}$ are to be
considered to be functions of $x$ (in the given case these functions are
constants), because a reference to $l_{i}$ as formal variables is absent.
After the transformation to another coordinate system the equation (\ref
{a6.2}) takes the form 
\begin{equation}
m\frac{d}{d\tau }\left[ \left( \tilde{l}_{k}\frac{d\tilde{q}}{d\tau }%
^{k}\right) ^{-1}\frac{d\tilde{q}^{i}}{d\tau }-\frac{1}{2}g^{ik}\tilde{l}%
_{k}\left( \tilde{l}_{j}\frac{d\tilde{q}^{j}}{d\tau }\right) ^{-2}\frac{d%
\tilde{q}}{d\tau }^{s}\frac{d\tilde{q}_{s}}{d\tau }\right] =e\tilde{F}%
^{il}\left( \tilde{q}\right) g_{lk}\frac{d\tilde{q}^{k}}{d\tau }
\label{a6.10}
\end{equation}

Here $\tilde{l}_{i}$ are considered as functions of $\tilde{x}$. But $\tilde{%
l}_{i}$ are other functions of $\tilde{x}$, than $l_{i}$ of $x$ (other
numerical constants $\tilde{l}_{k}=l_{j}\partial x^{j}/\partial \tilde{x}%
^{k} $ instead of $l_{k}$), and equations (\ref{a6.2}) and (\ref{a6.10})
have different forms with respect to quantities $q^{i}$, $F^{ik}$, $%
J^{i}\equiv 0$. It means that (\ref{a6.2}) is not relativistically covariant
with respect to $q^{i}$, $F^{ik}$, $J^{i}\equiv 0$, although it is
relativistically covariant with respect to $q^{i}$, $F^{ik}$, $l_{i}$, $%
J^{i}\equiv 0$.

Setting $l_{i}=\{1,0,0,0\}$, $t=q^{0}(\tau )$ in (\ref{a6.2}), we obtain 
\begin{equation}
m\frac{d^{2}q}{dt^{2}}^{\alpha }=eF_{.0}^{\alpha }+eF_{.\beta }^{\alpha }%
\frac{dq}{dt}^{\beta },\qquad i=\alpha =1,2,3;  \label{a6.11}
\end{equation}
\begin{equation}
\frac{m}{2}\frac{d}{dt}(\frac{dq}{dt}^{\alpha }\frac{dq}{dt}^{\alpha
})=eF_{.0}^{\alpha }\frac{dq}{dt}^{\alpha },\qquad i=0.  \label{a6.11a}
\end{equation}
Now the equations (\ref{a6.11}), (\ref{a6.11a}) do not contain the absolute
object $l_{k}$. It is easy to see that these equations describe a
nonrelativistic motion of a charged particle in a given electromagnetic
field $F^{ik}$. The fact that the equations (\ref{a6.2}) or (\ref{a6.11})
are nonrelativistic is connected with the space-time splitting into space
and time that is characteristic for Newtonian mechanics. Any 3-plane,
orthogonal to the vector $l^{k}$, consists of simultaneous events. This
space-time splitting is described in different ways in equations (\ref{a6.2}%
) and (\ref{a6.11}). It is described by the constant timelike vector $l_{k}$
in (\ref{a6.2}). In the equation (\ref{a6.11}) the space-time splitting is
described by a special choice of the coordinate system, whose time axis is
directed along the vector $l^{k}$, and all events having the same coordinate 
$t$ are simultaneous absolutely.

From physical viewpoint the Newtonian conception of space-time and that of
the special relativity distinguish in the number of invariants, describing
the space-time. In the Newtonian space-time conception there are two
independent invariant quantities: time $t$ and distance $r=\left| \mathbf{x}%
\right| $. This conception is associated with the description in terms of
the time $t$ and spatial coordinates $\mathbf{x}$. This description is known
as noncovariant description, and (\ref{a6.11}), (\ref{a6.11a}) is an example
of such a description. In the relativistic conception of the space-time
there is only one invariant quantity: space-time interval $s=\sqrt{%
c^{2}t^{2}-r^{2}}$. The relativistic conception of the space-time associated
with the description in terms of the 4-vector $x^{k}=\left\{ t,\mathbf{x}%
\right\} $. This description is known as (relativistically) covariant
description. Can we describe a relativistic phenomenon, using noncovariant
description, associated with the Newtonian conception? Yes, it is possible,
and noncovariant description of relativistic phenomena is used frequently.
Vice versa, can we describe a nonrelativistic phenomenon, using
relativistically covariant description, associated with the relativistic
conception of the space-time? Yes, it is possible. Relativistically
covariant description of the nonrelativistic phenomenon is possible,
provided this description contains two independent space-time invariants:
the time $t$ and the distance $r$. To obtain two invariants: $t$ and $r$
from the space-time interval $s$, one needs to introduce an absolute
constant timelike 4-vector $l^{k}$ ( $l^{k}l_{k}=1$). This vector is
associated with the space-time and admits one to construct two space-time
invariants $t$ and $r$ from the space-time interval $s$%
\[
t=x^{k}l_{k},\qquad r=\sqrt{\left( x^{k}l_{k}\right) ^{2}-x^{k}x_{k}},\qquad
c=1 
\]

Thus, any appearance of the absolute constant timelike 4-vector $l^{k}$ in
the relativistically covariant description of a physical phenomenon is open
to suspicion that the phenomenon is nonrelativistic (incompatible with the
relativity theory). The physical phenomenon is relativistic, if the vector $%
l^{k}$ is fictitious in the given description. On the contrary, the physical
phenomenon is nonrelativistic, because its description contains two
space-time invariants $t$ and $r$. The relativistically covariant
description is not used practically for description of nonrelativistic
phenomena, and most researchers are convinced, that any relativistically
covariant description is a description of a relativistic phenomenon.

The considered example shows that nonrelativistic equation (\ref{a6.11}) can
be written in a relativistically covariant form (\ref{a6.2}), provided one
introduces an absolute object $l_{i}$, describing space-time splitting.

Dirac matrices $\gamma ^{k}$ are absolute objects, as well as the metric
tensor $g^{kl}$, which may be considered as a derivative absolute object
determined by the relation (\ref{b1.2}). It follows from the relation (\ref
{b1.2}) that the $\gamma $-matrices are to be transformed as 4-vectors at
the linear transformations of the space-time coordinates $x$, because the
rhs of (\ref{b1.2}) is a tensor at the linear transformations of coordinates 
$x$. Although at the Lorentz transformation rhs of (\ref{b1.2}) is an
invariant, we cannot suppose that $\gamma ^{i}$ may be invariants, because $%
\gamma ^{i}$ cannot be invariants under linear transformations of
coordinates.

There are two approaches to the Dirac equation. In the first approach \cite
{S30,S51} the wave function $\psi $ is considered to be a scalar function
defined on the field of Clifford numbers $\gamma ^{l}$, 
\begin{equation}
\psi =\psi (x,\gamma )\Gamma ,\qquad \bar{\psi}=\Gamma \bar{\psi}(x,\gamma ),
\label{b1.4}
\end{equation}
where $\Gamma $ is a constant nilpotent factor which has the property $%
\Gamma F(\gamma )\Gamma =a\Gamma $. Here $F(\gamma )$ is arbitrary function
of $\gamma ^{l}$ and $a$ is a complex number, depending on the form of the
function $F$. Within such an approach $\psi $, $\bar{\psi}$ are transformed
as scalars under the Lorentz transformations, whereas $\gamma ^{l}$ are
transformed as components of a 4-vector. In this case the symmetry group of $%
\gamma ^{l}$ is a subgroup of the Lorentz group, and $\mathcal{S}_{\mathrm{D}%
}$ is a nonrelativistic dynamic system. Then the matrix vector $\gamma ^{l}$
describes some preferred direction in the space-time.

In the second (conventional) approach $\psi $ is considered to be a spinor,
and $\gamma ^{l}$, $l=0,1,2,3$ are scalars with respect to the
transformations of the Lorentz group. In this case the symmetry group of the
absolute objects $\gamma ^{l}$ is the Lorentz group, and dynamic system $%
\mathcal{S}_{\mathrm{D}}$ is considered to be a relativistic dynamic system.

Of course, the approaches leading to incompatible conclusions cannot be both
valid. At least, one of them is wrong. Analyzing the two approaches,
Sommerfeld \cite{S51} considered the first approach to be more reasonable.
In the second case the analysis is rather difficult due to non-standard
transformations of $\gamma ^{l}$ and $\psi $ under linear coordinate
transformations $T$. Indeed, the transformation $T$ for the vector $j^{l}=%
\bar{\psi}\gamma ^{l}\psi $ has the form 
\begin{equation}
\tilde{\overline{\psi }}\tilde{\gamma}^{l}\tilde{\psi}=\frac{\partial \tilde{%
x}^{l}}{\partial x^{s}}\bar{\psi}\gamma ^{s}\psi ,  \label{b1.5}
\end{equation}
where the quantities marked by tilde mean the quantities in the transformed
coordinate system. This transformation can be carried out by two different
ways 
\begin{equation}
1:\;\;\;\tilde{\psi}=\psi ,\qquad \tilde{\overline{\psi }}=\overline{\psi }%
,\qquad \tilde{\gamma}^{l}=\frac{\partial \tilde{x}^{l}}{\partial x^{s}}%
\gamma ^{s},\qquad l=0,1,2,3  \label{b1.6}
\end{equation}
\begin{equation}
2:\;\;\;\tilde{\gamma}^{l}=\gamma ^{l},\qquad l=0,1,2,3,\qquad \tilde{\psi}%
=S(\gamma ,T)\psi ,\qquad \tilde{\overline{\psi }}=\overline{\psi }%
S^{-1}(\gamma ,T),  \label{b1.7}
\end{equation}
\begin{equation}
S^{\ast }(\gamma ,T)\gamma ^{0}=\gamma ^{0}S^{-1}(\gamma ,T)  \label{b1.8b}
\end{equation}
The relations (\ref{b1.6}) correspond to the first approach and the
relations (\ref{b1.7}) correspond to the second one. Both ways (\ref{b1.6})
and (\ref{b1.7}) lead to the same result, provided 
\begin{equation}
S^{-1}(\gamma ,T)\gamma ^{l}S(\gamma ,T)=\frac{\partial \tilde{x}^{l}}{%
\partial x^{s}}\gamma ^{s}  \label{b1.9}
\end{equation}
In particular, for the infinitesimal Lorentz transformation $%
x^{i}\rightarrow x^{i}+\delta \omega _{.k}^{i}x^{k}$ the matrix $S(\gamma
,T) $ has the form \cite{S61} 
\begin{equation}
S(\gamma ,T)=\exp \left( \frac{\delta \omega _{ik}}{8}\left( \gamma
^{i}\gamma ^{k}-\gamma ^{k}\gamma ^{i}\right) \right)  \label{b1.10}
\end{equation}
The second way (\ref{b1.7}) has two defects. First, the transformation law
of $\psi $ depends on $\gamma $, i.e. under linear transformation $T$ of
coordinates the components of $\psi $ are transformed through $\psi $ and $%
\gamma ^{l}$, but not only through $\psi $. Note that the tensor components
in a coordinate system are transformed only through the tensor components in
other coordinate system, and this transformation does not contain any
absolute objects. (for instance, the relation (\ref{b1.5})). Second, the
relation (\ref{b1.9}) is compatible with (\ref{b1.2}) only under
transformations $T$ between orthogonal coordinate systems, when components $%
g^{lk}=\{1,-1,-1,-1\}$ of the metric tensor are invariant. Indeed, lhs of
the relation (\ref{b1.2}) is a scalar under any linear coordinate
transformations, whereas rhs of (\ref{b1.2}) is invariant only under
orthogonal (Lorentz) transformations. Rhs of the relation (\ref{b1.2}) is
transformed as a tensor under linear coordinate transformations. In other
words, at the second approach the relation (\ref{b1.2}) is not covariant, in
general, with respect to arbitrary linear transformations of coordinates. In
this case one cannot be sure that the symmetry group of the dynamic system
coincides with the symmetry group of absolute objects.

The fact that the symmetry group of the dynamic system coincides with the
symmetry group of absolute objects was derived with the supposition, that
under the coordinate transformation any object is transformed only via its
components. This condition is violated in the second case, and one cannot be
sure that the symmetry group of dynamic system coincides with that of
absolute objects.

After change of variables the action (\ref{b1.1}) is transformed to the form
(\ref{c4.15}) -- (\ref{c4.18}), the $\gamma $-matrices being eliminated. But
after reduction of the action to the relativistically covariant form two new
absolute objects appear: constant 4-vector $f^{k}$ and constant
4-pseudovector $z^{i}$. The 4-pseudovector $z^{i}$ appears to be fictitious.
But the action depends really on 4-vector $f^{k}$, which resembles the
vector $l_{k}$ in the considered example (\ref{a6.2}). It means that the
dynamic system $\mathcal{S}_{\mathrm{D}}$ is nonrelativistic, because it
supposes an absolute separation of the space-time into the space and the
time.

\section{Modification of the classical Dirac particle}

Not all terms in the action (\ref{c4.15a}) -- (\ref{c4.18a}) contain
absolute objects $f^{k}$ and $z^{k}$. Two first terms of (\ref{c4.16a}) and
the term (\ref{c4.17a}) do not contain absolute objects. It means that the
Dirac particle $\mathcal{S}_{\mathrm{D}}$ is described partly
relativistically and partly nonrelativistically. To determine what is
described relativistically, it is useful to investigate the classical Dirac
particle $\mathcal{S}_{\mathrm{Dcl}}$, which is a discrete dynamic system,
i.e. the dynamic system having a finite number of the freedom degrees. The
classical Dirac particle $\mathcal{S}_{\mathrm{Dcl}}$ is determined uniquely
by the Dirac particle $\mathcal{S}_{\mathrm{D}}$ by means of the
relativistic procedure (\ref{b1.8}). Investigating the classical Dirac
particle $\mathcal{S}_{\mathrm{Dcl}}$, we can determine which degrees of
freedom are described nonrelativistically and correct the nonrelativistic
description.

After dynamic disquantization (\ref{b1.8}) of the action (\ref{c4.15}) we
obtain the action for the classical Dirac particle $\mathcal{S}_{\mathrm{Dcl}%
}$. According to calculations of \cite{R2004} we obtain 
\begin{equation}
\mathcal{S}_{\mathrm{Dcl}}:\qquad \mathcal{A}_{\mathrm{Dcl}}[x,\mathbf{\xi }%
]=\int \left\{ -\kappa _{0}m\sqrt{\dot{x}^{i}\dot{x}_{i}}+\hbar {\frac{(\dot{%
\mathbf{\xi }}\times \mathbf{\xi })\mathbf{z}}{2(1+\mathbf{\xi z})}}+\hbar 
\frac{(\dot{\mathbf{x}}\times \ddot{\mathbf{x}})\mathbf{\xi }}{2\sqrt{\dot{x}%
^{s}\dot{x}_{s}}(\sqrt{\dot{x}^{s}\dot{x}_{s}}+\dot{x}^{0})}\right\} d\tau
_{0}  \label{b3.9}
\end{equation}
where coordinates $x=x\left( \tau _{0}\right) =\left\{ x^{0},\mathbf{x}%
\right\} =\left\{ x^{0},x^{2},x^{2},x^{3}\right\} $ of the classical Dirac
particle $\mathcal{S}_{\mathrm{Dcl}}$ and its internal variables $\mathbf{%
\xi =\xi }\left( \tau _{0}\right) \mathbf{=}\left\{ \xi _{1},\xi _{2},\xi
_{3}\right\} $, $\mathbf{\xi }^{2}=1$ are considered to be functions of $%
\tau _{0}$. The quantity $\kappa _{0}=\pm 1$ is obtained as a result of
solution \cite{R2004} of dynamic equation for the dynamic variable $\kappa $.

The action (\ref{b3.9}) can be written in the relativistically covariant
form 
\begin{equation}
\mathcal{S}_{\mathrm{Dcl}}:\qquad \mathrm{\mathcal{A}}_{\mathrm{Dcl}}\left[
x,\xi \right] =\int \{-\kappa _{0}m\sqrt{\dot{x}^{i}\dot{x}_{i}}-\hbar \frac{%
\varepsilon _{iklm}\xi ^{i}\dot{\xi}^{k}f^{l}z^{m}}{2(1-\xi ^{s}z_{s})}+%
\frac{\hbar }{2}Q\varepsilon _{iklm}\dot{x}^{i}\ddot{x}^{k}f^{l}\xi
^{m}\}d\tau _{0}  \label{f5.1}
\end{equation}
\begin{equation}
Q=Q\left( \dot{x},f\right) =\frac{1}{\sqrt{\dot{x}^{s}\dot{x}_{s}}(\dot{x}%
^{l}f_{l}+\sqrt{\dot{x}^{l}\dot{x}_{l}})},\qquad \dot{x}^{l}\equiv \frac{%
dx^{l}}{d\tau _{0}}  \label{f5.2}
\end{equation}
where 4-vectors $f^{k}$, $z^{k}$ are defined respectively by relations (\ref
{d2.1}), (\ref{d2.8}) and 
\begin{equation}
\xi ^{k}=\left\{ \xi _{0},\mathbf{\xi }\right\} ,\qquad \xi ^{l}f_{l}=0
\label{f5.2a}
\end{equation}
The 4-vector $z^{k}$ appears to be fictitious (see Appendix B of this paper
or of the paper \cite{R2004}). But the 4-vector $f^{k}$ is not fictitious.

The action (\ref{f5.1}) is invariant with respect to a change of independent
variable $\tau _{0}$%
\begin{equation}
\tau _{0}\rightarrow \tilde{\tau}_{0}=F\left( \tau _{0}\right)  \label{f5.2b}
\end{equation}
where $F$ is an arbitrary monotone function. The variable $\tau _{0}$ may be
chosen in such a way that 
\begin{equation}
\sqrt{\dot{x}^{s}\dot{x}_{s}}=1  \label{f5.2c}
\end{equation}
for all values of the independent variable $\tau _{0}$. Formally the
expression (\ref{f5.2c}) appears to be an integral of dynamic equations
generated by the action (\ref{f5.1}).

The first term in (\ref{f5.1}) is described relativistically, because it
does not contain the absolute object $f^{k}$. This term describes the
translation degrees of freedom. Two last terms contain the absolute object $%
f^{k}$. They describe internal degrees of freedom. Description of internal
degrees of freedom appears to be nonrelativistic.

If we set $\hbar =0$ in the action (\ref{f5.1}), we suppress internal
degrees of freedom of the classical Dirac particle, and description of $%
\mathcal{S}_{\mathrm{Dcl}}$ becomes relativistical. The classical Dirac
particle becomes relativistic, provided the absolute object $f^{k}$ is
changed by the dynamical variables. For instance, we may identify the unit
timelike 4-vector $f^{k}$ with the constant energy-momentum vector $p_{k}$,
normalized in a proper way. At such a change the classical Dirac particle $%
\mathcal{S}_{\mathrm{Dcl}}$ turns into the modified classical Dirac particle 
$\mathcal{S}_{\mathrm{mDcl}}$.

To make this change we introduce designations 
\begin{equation}
y^{l}=\dot{x}^{l},\qquad l=0,1,2,3  \label{f5.3}
\end{equation}
and rewrite the action (\ref{f5.1}) in the form 
\begin{equation}
\mathcal{S}_{\mathrm{Dcl}}:\;\;\mathrm{\mathcal{A}}_{\mathrm{Dcl}}\left[
x,y,\xi ,p\right] =\int \left\{ L\left( y,\dot{y},\xi ,\dot{\xi},f\right)
-p_{l}\left( y^{l}-\dot{x}^{l}\right) \right\} d\tau _{0}  \label{f5.4}
\end{equation}
where 
\begin{equation}
L\left( y,\dot{y},\xi ,\dot{\xi},f\right) =-\kappa _{0}\sqrt{y_{l}y^{l}}%
-\hbar \frac{\varepsilon _{iklm}\xi ^{i}\dot{\xi}^{k}f^{l}z^{m}}{2(1-\xi
^{s}z_{s})}+\frac{\hbar }{2}Q\varepsilon _{iklm}y^{i}\dot{y}^{k}f^{l}\xi ^{m}
\label{f5.6}
\end{equation}
\begin{equation}
Q=Q\left( y,f\right) =\frac{1}{\sqrt{y^{s}y_{s}}(y^{l}f_{l}+\sqrt{y^{l}y_{l}}%
)},  \label{f5.7}
\end{equation}
Variables $p_{l}=p_{l}\left( \tau _{0}\right) $, $l=0,1,2,3$ are the
Lagrange multipliers, introducing designations (\ref{f5.3}). The quantities $%
f^{k}$ and $z^{k}$ are not dynamical variables and they are not to be varied.

Dynamic equations generated by the action (\ref{f5.4}) have the form 
\begin{eqnarray}
\frac{\delta \mathcal{A}_{\mathrm{Dcl}}}{\delta x^{l}} &=&-\dot{p}%
_{l}=0,\qquad p_{l}=\text{const}  \label{f5.8a} \\
\frac{\delta \mathcal{A}_{\mathrm{Dcl}}}{\delta y^{l}} &=&\frac{\partial L}{%
\partial y^{l}}-\frac{d}{d\tau _{0}}\frac{\partial L}{\partial \dot{y}^{l}}%
-p_{l}=0,  \label{f5.9} \\
\frac{\delta \mathcal{A}_{\mathrm{Dcl}}}{\delta \xi ^{l}} &=&\left( \frac{%
\partial L}{\partial \xi ^{s}}-\frac{d}{d\tau _{0}}\frac{\partial L}{%
\partial \dot{\xi}^{s}}\right) \left( \delta _{l}^{s}-f_{l}f^{s}+\xi _{l}\xi
^{s}\right) =0  \label{f5.11} \\
\frac{\delta \mathcal{A}_{\mathrm{Dcl}}}{\delta p_{l}} &=&\dot{x}^{l}-y^{l}=0
\label{f5.12a}
\end{eqnarray}
The second multiplier in (\ref{f5.11}) takes into account that the
pseudovector $\xi ^{k}$ restricted by the constraints $\xi _{k}\xi ^{k}=-1$, 
$\xi _{k}f^{k}=0$.

As far as $p_{l}=$const, $l=0,1,2,3$ in force of dynamic equations, we can
express the constant unit 4-vector $f_{l}$ in the form 
\begin{equation}
f_{l}=\frac{\varepsilon p_{l}}{\sqrt{p_{s}p^{s}}}\qquad l=0,1,2,3,\qquad
\varepsilon =\text{sgn}p_{0}=\pm 1  \label{f5.18}
\end{equation}
and substitute $f_{l}$ from (\ref{f5.18}) in the action (\ref{f5.4}). We
obtain 
\begin{equation}
\mathcal{S}_{\mathrm{mDcl}}:\qquad \mathrm{\mathcal{A}}_{\mathrm{mDcl}}\left[
x,y,\xi ,p\right] =\int \left\{ L\left( y,\dot{y},\xi \mathbf{,}\dot{\xi}%
\mathbf{,}\frac{\varepsilon p_{l}}{\sqrt{p_{s}p^{s}}}\right) +p_{l}\left( 
\dot{x}^{l}-y^{l}\right) \right\} d\tau _{0}  \label{f5.19}
\end{equation}

The action (\ref{f5.19}) describes dynamic system $\mathcal{S}_{\mathrm{mDcl}%
}$, which distinguishes, in general, from the dynamic system $\mathcal{S}_{%
\mathrm{Dcl}}$. The dynamic system $\mathcal{S}_{\mathrm{mDcl}}$ is
compatible with the relativity principles, whereas the dynamic system $%
\mathcal{S}_{\mathrm{Dcl}}$ is not. In particular, it means as follows. Let $%
\left\{ x,y,\xi ,p\right\} $ be a solution of dynamic equations, generated
by the action (\ref{f5.19}) and $\left\{ \tilde{x},\tilde{y},\tilde{\xi},%
\tilde{p}\right\} $ have been obtained from $\left\{ x,y,\xi ,p\right\} $ by
means of some transformation of the Lorentz group. Then $\left\{ \tilde{x},%
\tilde{y},\tilde{\xi},\tilde{p}\right\} $ is a solution of the same dynamic
equations. Solutions of dynamic equations (\ref{f5.8a}) -- (\ref{f5.12a})
have not these property, in general.

All dynamic equations, generated by the action (\ref{f5.19}) coincide with
the dynamic equations (\ref{f5.8a}) -- (\ref{f5.12a}) except for the dynamic
equation (\ref{f5.12a}), which has now the form 
\begin{equation}
\frac{\delta \mathcal{A}_{\mathrm{mDcl}}}{\delta p_{l}}=-y^{l}+\dot{x}^{l}+%
\frac{\varepsilon }{\sqrt{p_{s}p^{s}}}\left[ \frac{\partial L\left( y,\dot{y}%
,\xi ,\dot{\xi},f\right) }{\partial f_{k}}\right] _{f_{i}=\frac{\varepsilon
p_{i}}{\sqrt{p_{s}p^{s}}}}\left( \delta _{k}^{l}-\frac{p_{k}p^{l}}{p_{s}p^{s}%
}\right) =0  \label{f5.20}
\end{equation}

We write dynamic equations for the action (\ref{f5.19}) in the coordinate
system, where the 4-vector $f_{l}$ and the energy-momentum 4-vector $p_{l}$
have the form 
\begin{equation}
p_{l}=\left\{ p_{0},0,0,0\right\} ,\qquad f_{l}=\left\{ 1,0,0,0\right\}
\label{f5.21}
\end{equation}
Solutions for other values of the canonical momentum $p_{l}=\tilde{p}_{l}$
may be obtained from solutions for $p_{l}=\left\{ p_{0},0,0,0\right\} $ by
means of the coordinate transformation of the Lorentz group, which
transforms $p_{l}=\left\{ p_{0},0,0,0\right\} $ into $p_{l}=\tilde{p}_{l}$, $%
l=0,1,2,3$.

Dynamic equations (\ref{f5.8a}) -- (\ref{f5.11}) generated by the action (%
\ref{f5.4}) as well as by the action (\ref{f5.19}) do not depend on the
variables $x^{l}$. They can be solved independently of the solution of the
equation (\ref{f5.12a}), or (\ref{f5.20}). In the paper \cite{R2004} these
equations have been solved in the coordinate system, where the Dirac
particle is at rest and the energy-momentum vector $p_{l}$ has the form (\ref
{f5.21}). For the sake of convenience this solution is presented in the
Appendix A.

The solution (\ref{e6.21}) has the form 
\begin{eqnarray}
p_{k} &=&\left\{ -\frac{\kappa _{0}m}{\gamma },0,0,0\right\}  \label{f5.22}
\\
\zeta ^{l} &=&\left\{ 0,z^{1},z^{2},z^{3}\right\} ,\qquad \xi ^{l}=\left\{
0,0,0,\varepsilon _{0}\right\} ,\qquad \varepsilon _{0}=\pm 1  \label{f5.22a}
\end{eqnarray}
\begin{eqnarray}
y^{k} &=&y^{k}\left( \tau _{0}\right) =\left\{ \gamma ,\sqrt{\gamma ^{2}-1}%
\cos \Phi ,\sqrt{\gamma ^{2}-1}\sin \Phi ,0\right\}  \label{f5.23} \\
\dot{y}^{k} &=&\left\{ 0,-\omega \sqrt{\gamma ^{2}-1}\sin \Phi ,\omega \sqrt{%
\gamma ^{2}-1}\cos \Phi ,0\right\}  \label{f5.24}
\end{eqnarray}
\begin{equation}
\Phi =-\frac{2\varepsilon _{0}\kappa _{0}m}{\hbar \gamma }\tau _{0}+\phi
,\qquad \omega =\frac{d\Phi }{d\tau _{0}}=-\frac{2\varepsilon _{0}\kappa
_{0}m}{\hbar \gamma }  \label{f5.24a}
\end{equation}
where $\phi $ and $\gamma $ ($\gamma ^{2}\geq 1)$ are arbitrary constants.

Substituting relations (\ref{f5.22}) - (\ref{f5.24}) in the equation (\ref
{f5.20}), we can express the variables $\dot{x}^{k}$ as functions of $\tau
_{0}$. These equations can be integrated easily.

From the relations (\ref{f5.6}), (\ref{f5.7}) we obtain the following
expression 
\begin{equation}
\frac{\partial L}{\partial f_{l}}=-\hbar g^{lp}\frac{\varepsilon _{pikm}\xi
^{i}\dot{\xi}^{k}z^{m}}{2(1-\xi ^{s}z_{s})}+\frac{\hbar }{2}g^{lp}\frac{%
\varepsilon _{pikm}y^{i}\dot{y}^{k}\xi ^{m}}{\sqrt{y^{s}y_{s}}(y^{s}f_{s}+%
\sqrt{y^{l}y_{l}})}-\frac{\hbar }{2}\frac{y^{l}\varepsilon _{iksm}y^{i}\dot{y%
}^{k}f^{s}\xi ^{m}}{\sqrt{y^{s}y_{s}}(y^{s}f_{s}+\sqrt{y^{l}y_{l}})^{2}}
\label{f5.25}
\end{equation}
The action (\ref{f5.19}) as well as the original action (\ref{f5.4}) are
invariant with respect to transformation of the independent variable $\tau
_{0}$ 
\[
\tau _{0}\rightarrow \tilde{\tau}_{0}=F\left( \tau _{0}\right) \qquad
y^{l}\rightarrow \tilde{y}^{l}=y^{l}\left( \frac{dF\left( \tau _{0}\right) }{%
d\tau _{0}}\right) ^{-1},\qquad \xi ^{l}\rightarrow \xi ^{l},\qquad
p_{l}\rightarrow p_{l},\qquad x^{l}\rightarrow x^{l} 
\]
and we choose the independent variable $\tau _{0}$ in such a way, that 
\begin{equation}
y_{s}y^{s}=1  \label{f5.26}
\end{equation}
For the dynamic system $\mathcal{S}_{\mathrm{Dcl}}$ the condition (\ref
{f5.26}) is equivalent to the condition 
\begin{equation}
\dot{x}_{s}\dot{x}^{s}=1,  \label{f5.26a}
\end{equation}
because in this case $\dot{x}^{l}=y^{l}$. But for the dynamic system $%
\mathcal{S}_{\mathrm{mDcl}}$ the conditions (\ref{f5.26}) and (\ref{f5.26a})
are not equivalent, because in this case $\dot{x}^{l}\neq y^{l}$, in
general, as it follows from (\ref{f5.20}).

Under constraints (\ref{f5.22}), (\ref{f5.26}) the additional term of the
modified equation (\ref{f5.20}) has the form 
\begin{equation}
\left[ \frac{\partial L\left( y,\dot{y},\xi ,\dot{\xi},f\right) }{\partial
f_{k}}\right] _{f_{l}=\frac{\varepsilon p_{l}}{\sqrt{p_{i}p^{i}}}}\left(
\delta _{k}^{l}-\frac{p_{k}p^{l}}{p_{s}p^{s}}\right) =\left\{ 
\begin{array}{l}
0,\;\;\;\text{ if\ \ }l=0 \\ 
\frac{\varepsilon }{\left| p_{0}\right| }\frac{\partial L}{\partial f_{\mu }}%
,\text{\ \ \ if \ \ }l=\mu =1,2,3
\end{array}
\right.  \label{f5.27}
\end{equation}
\begin{eqnarray}
\frac{\partial L}{\partial f_{\mu }} &=&-\frac{\hbar }{2}\frac{\varepsilon
_{\mu 0\alpha \beta }y^{0}\dot{y}^{\alpha }\xi ^{\beta }}{(y^{0}+1)}+\frac{%
\hbar }{2}\frac{\varepsilon _{\mu 0\alpha \beta }\dot{y}^{0}y^{\alpha }\xi
^{\beta }}{(y^{0}+1)}-\frac{\hbar }{2}\frac{y^{\mu }\varepsilon _{\alpha
\beta \gamma }y^{\alpha }\dot{y}^{\beta }\xi ^{\gamma }}{(y^{0}+1))^{2}} 
\nonumber \\
&=&\frac{\hbar }{2(y^{0}+1)}\left( -\varepsilon _{\mu \alpha \beta }y^{0}%
\dot{y}^{\alpha }\xi ^{\beta }+\varepsilon _{\mu \alpha \beta }\dot{y}%
^{0}y^{\alpha }\xi ^{\beta }+\frac{y^{\mu }\varepsilon _{\alpha \beta \gamma
}y^{\alpha }\dot{y}^{\beta }\xi ^{\gamma }}{(y^{0}+1)}\right) ,
\label{f5.28} \\
\mu &=&1,2,3  \nonumber
\end{eqnarray}
where $\varepsilon _{\alpha \beta \gamma }$ is the Levi-Chivita pseudotensor
in the three-dimensional space. Let us substitute (\ref{f5.27}), (\ref{f5.28}%
) in the dynamic equation (\ref{f5.20}). We obtain after calculation 
\begin{equation}
\dot{x}^{0}=\gamma \text{,\qquad }\dot{x}^{1}=\gamma \sqrt{\frac{\gamma -1}{%
\gamma +1}}\cos \Phi ,\qquad \dot{x}^{2}=\gamma \sqrt{\frac{\gamma -1}{%
\gamma +1}}\sin \Phi ,\qquad \dot{x}^{3}=0  \label{f5.32}
\end{equation}
where $\Phi $ is determined by the relation (\ref{f5.24a}).

Instead of (\ref{f5.26a}) we obtain 
\begin{equation}
\dot{x}^{l}\dot{x}_{l}=\gamma ^{2}-\left( \gamma ^{2}-1\right) \left( \frac{%
\gamma }{\gamma +1}\right) ^{2}=2\frac{\gamma ^{2}}{\gamma +1}  \label{f5.33}
\end{equation}

Let now $t=x^{0}$ be the independent variable instead of $\tau _{0}$. Then
equations (\ref{f5.32}) are transformed into 
\begin{equation}
\frac{dx^{k}}{dt}=\left\{ 1,\sqrt{\frac{\gamma -1}{\gamma +1}}\cos \Phi
,-\varepsilon _{0}\kappa _{0}\sqrt{\frac{\gamma -1}{\gamma +1}}\sin \Phi
,0\right\}  \label{f5.34}
\end{equation}
where 
\begin{equation}
\Phi =\Omega t-\varepsilon _{0}\kappa _{0}\phi ,\qquad \Omega =\frac{2m}{%
\hbar \gamma ^{2}}  \label{f5.35}
\end{equation}
Integration of (\ref{f5.34}) leads to the relations 
\begin{equation}
x^{k}=\left\{ t,\frac{\hbar \gamma ^{2}}{2m}\sqrt{\frac{\gamma -1}{\gamma +1}%
}\sin \Phi ,\frac{\hbar \varepsilon _{0}\kappa _{0}\gamma ^{2}}{2m}\sqrt{%
\frac{\gamma -1}{\gamma +1}}\cos \Phi ,0\right\}  \label{f5.36}
\end{equation}
which describe the world line of the modified classical Dirac particle $%
\mathcal{S}_{\mathrm{mDcl}}$. The total mass $M_{\mathrm{mDcl}}$ of $%
\mathcal{S}_{\mathrm{mDcl}}$ is described by the relation 
\begin{equation}
M_{\mathrm{mDcl}}=\sqrt{p_{0}^{2}-\mathbf{p}^{2}}=\left| p_{0}\right| =\frac{%
m}{\gamma },\qquad \gamma \geq 1  \label{f5.37}
\end{equation}
World line of $\mathcal{S}_{\mathrm{mDcl}}$ is a helix with the radius 
\begin{equation}
a_{\mathrm{mDcl}}=\frac{\hbar \gamma ^{2}}{2m}\sqrt{\frac{\gamma -1}{\gamma
+1}}  \label{f5.38}
\end{equation}
The constant $\gamma \geq 1$ describes the intensity of excitation of the
internal degrees of freedom.

For the classical Dirac particle $\mathcal{S}_{\mathrm{Dcl}}$ with
nonrelativistic description of the internal degrees of freedom the relations
(\ref{f5.36}) - (\ref{f5.38}) have another form \cite{R2004} 
\begin{equation}
x_{\mathrm{Dcl}}^{k}=\left\{ t,\frac{\hbar \gamma }{2m}\sqrt{\gamma ^{2}-1}%
\sin \left( \Omega _{\mathrm{Dcl}}t\right) ,\frac{\hbar \gamma }{2m}\sqrt{%
\gamma ^{2}-1}\cos \left( \Omega _{\mathrm{Dcl}}t\right) ,0\right\}
\label{f5.39}
\end{equation}
\begin{equation}
M=M_{\mathrm{Dcl}}=M_{\mathrm{mDcl}}=\frac{m}{\gamma },\qquad \Omega =\Omega
_{\mathrm{Dcl}}=\Omega _{\mathrm{mDcl}}=\frac{2m}{\hbar \gamma ^{2}},\qquad
a_{\mathrm{Dcl}}=\frac{\hbar \gamma }{2m}\sqrt{\gamma ^{2}-1}  \label{f5.40}
\end{equation}

The dynamical systems $\mathcal{S}_{\mathrm{mDcl}}$ and $\mathcal{S}_{%
\mathrm{Dcl}}$ distinguish only in the radius $a$ of the helix 
\begin{equation}
a_{\mathrm{mDcl}}=\frac{\gamma }{\gamma +1}a_{\mathrm{Dcl}}=\frac{\hbar
\gamma ^{2}}{2m}\sqrt{\frac{\gamma -1}{\gamma +1}}  \label{f5.41}
\end{equation}
The ratio between $a_{\mathrm{mDcl}}$ and $a_{\mathrm{Dcl}}$ is maximal in
the case of the slight excitation, when $\gamma =1$, whereas the difference 
\begin{equation}
a_{\mathrm{Dcl}}-a_{\mathrm{mDcl}}=\frac{\hbar }{2m}\gamma \sqrt{\frac{%
\gamma -1}{\gamma +1}}  \label{f5.42}
\end{equation}
has minimum at $\gamma =1$ and no maximum.

Interpretation of the modified classical Dirac particle $\mathcal{S}_{%
\mathrm{mDcl}}$ is the same as the interpretation of $\mathcal{S}_{\mathrm{%
Dcl}}$ \cite{R2004}. The dynamic system $\mathcal{S}_{\mathrm{mDcl}}$ is
interpreted as a relativistic rotator (two coupled particles of the mass $%
m_{0}$, rotating around their mass center). Coordinates $x_{\mathrm{(1)}}$, $%
x_{\mathrm{(2)}}$ of the rotator constituents in the center mass coordinate
system are described by the relations 
\begin{eqnarray}
x_{\mathrm{(1)}}^{k} &=&\left\{ t,a\sin \left( \Omega t+\phi \right) ,a\sin
\left( \Omega t+\phi \right) ,0\right\}  \label{f5.43} \\
x_{\mathrm{(2)}}^{k} &=&\left\{ t,-a\sin \left( \Omega t+\phi \right)
,-a\sin \left( \Omega t+\phi \right) ,0\right\}  \label{f5.44}
\end{eqnarray}
where $a$ , $M$ and $\Omega $ are respectively the radius of the rotator
helix, the total mass and the frequency determined by the relations (\ref
{f5.41}), (\ref{f5.40}). The quantities $a$, $M$, $\Omega $ are connected by
the relation 
\begin{equation}
M=\frac{2m_{0}}{\sqrt{1-a^{2}\Omega ^{2}}}  \label{f5.45}
\end{equation}

The main characteristic of the relativistic rotator is the rigidity function 
$f_{\mathrm{r}}$, describing rigidity of coupling between the rotator
constituents. It is defined by the relation 
\begin{equation}
f_{\mathrm{r}}=f_{\mathrm{r}}\left( a\right) =\frac{M-2m_{0}}{2m_{0}}=\frac{1%
}{\sqrt{1-a^{2}\Omega ^{2}}}-1  \label{f5.46}
\end{equation}
where $f_{\mathrm{r}}$ is considered as a function of the radius $a$. Giving
the rigidity function as a function of $a$ , we determine by means of (\ref
{f5.46}) the relation between $\Omega $ and $a$. As a result we obtain that
the quantities $a$, $M$, $\Omega $ are functions of the mass $m_{0}$, and of
some parameter $\gamma $, describing the state of the rotator. Relations (%
\ref{f5.40}), (\ref{f5.41}) determine the quantities $a$, $M$, $\Omega $ of
the modified Dirac particle $\mathcal{S}_{\mathrm{mDcl}}$ as functions of
the parameter $\gamma $, describing the state of $\mathcal{S}_{\mathrm{mDcl}%
} $, and of the mass $m$, which is a parameter of $\mathcal{S}_{\mathrm{mDcl}%
}$. Identification of $\mathcal{S}_{\mathrm{mDcl}}$ with the relativistic
rotator is possible, provided we assume that the relation between parameter $%
m$ of $\mathcal{S}_{\mathrm{mDcl}}$ and the rotator parameter $m_{0}$
depends on the state $\gamma $ of dynamic system $\mathcal{S}_{\mathrm{mDcl}%
} $.

Identifying the rotator with $\mathcal{S}_{\mathrm{mDcl}}$, we are forced to
choose, what of quantities $m$ or $m_{0}$ is a parameter of $\mathcal{S}_{%
\mathrm{mDcl}}$. From physical viewpoint it seems that $m_{0}$ is to be the
real parameter of $\mathcal{S}_{\mathrm{mDcl}}$. At such a choice the Dirac
mass $m$, as well as the total mass $M$ are functions of the rotation state,
described by the parameter $\gamma $, which is an integral of motion of $%
\mathcal{S}_{\mathrm{mDcl}}$. It means that the Dirac mass $m$ as well as
the quantities $m$, $\Omega $, $a$, $M$ are integrals of motion of the
rotator, whereas $m_{0}$ is a parameter of the dynamic system.
Unfortunately, such an identification disagrees with the definition of $%
\mathcal{S}_{\mathrm{mDcl}}$ as a dynamic system described by the action (%
\ref{f5.19}). On the other hand, if we consider the Dirac mass $m$ as a
parameter of $\mathcal{S}_{\mathrm{mDcl}}$ (but not as an integral of
motion), the situation, described by the relations (\ref{f5.40}), is such,
that the total mass $M=m/\gamma $ decreases, whereas the internal velocity $%
\left( \gamma -1\right) ^{1/2}\left( \gamma +1\right) ^{-1/2}$ increases
with increasing $\gamma $. It means that the mass $m_{0}$ of the kinetic
energy bearer decreases with increasing $\gamma $.

Eliminating $M,\Omega $ and $a$ from relations (\ref{f5.45}), (\ref{f5.41})
and (\ref{f5.40}), we obtain the relation between $m$ and $m_{0}$ for $%
\mathcal{S}_{\mathrm{mDcl}}$ and for $\mathcal{S}_{\mathrm{Dcl}}$ 
\begin{equation}
\mathcal{S}_{\mathrm{mDcl}}:\qquad m=\sqrt{2\gamma \left( \gamma +1\right) }%
m_{0}  \label{f5.47}
\end{equation}
\begin{equation}
\mathcal{S}_{\mathrm{Dcl}}:\qquad m=2m_{0}\gamma ^{2}  \label{f5.48}
\end{equation}
The rigidity function for $\mathcal{S}_{\mathrm{Dcl}}$ has the form 
\begin{equation}
\mathcal{S}_{\mathrm{Dcl}}:\qquad f_{\mathrm{r}}\left( a\right) =\frac{1}{%
\sqrt{1-\left( \frac{4am_{0}}{\hbar }\right) ^{2}}}-1  \label{f5.49}
\end{equation}
The rigidity function $f_{\mathrm{r}}\left( a\right) $ for $\mathcal{S}_{%
\mathrm{mDcl}}$ is given implicitly by the relation 
\begin{equation}
\mathcal{S}_{\mathrm{mDcl}}:\qquad a=\frac{\hbar }{4m_{0}}\frac{\left(
2\left( f_{\mathrm{r}}+1\right) ^{2}-1\right) ^{3/2}}{\sqrt{2}\left( f_{%
\mathrm{r}}+1\right) ^{3}}\sqrt{\left( f_{\mathrm{r}}+1\right) ^{2}-1}
\label{f5.50}
\end{equation}
To compare the rigidity functions for $\mathcal{S}_{\mathrm{Dcl}}$ and $%
\mathcal{S}_{\mathrm{mDcl}}$ we write the relation (\ref{f5.49}) in the form
resolved with respect to $a$. We obtain 
\begin{equation}
\mathcal{S}_{\mathrm{Dcl}}:\qquad a=\frac{\hbar }{4m_{0}}\sqrt{1-\frac{1}{%
\left( f_{\mathrm{r}}+1\right) ^{2}}}  \label{f5.51}
\end{equation}

Comparison of (\ref{f5.50}) and (\ref{f5.51}) shows, that both rigidity
functions are close at small $a$ $\;\;(0<4m_{0}a/\hbar <0.4)$. For large $a$
the coupling, described by the rigidity function $f_{\mathrm{r}}$ is more
rigid for $\mathcal{S}_{\mathrm{Dcl}}$, than for $\mathcal{S}_{\mathrm{mDcl}%
} $.

\section{Discussion}

Almost eighty years since its appearance the Dirac equation exemplified the
most useful relativistic dynamic equation. Hyperfine structure of the
hydrogen spectrum has been explained by means of the Dirac equation. The
quantum electrodynamics is based on the Dirac equation. How can the newly
discovered internal degrees of freedom of the Dirac particle influence on
these well established results?

Only internal degrees of freedom of the Dirac particle are described
incorrectly (nonrelativistically), whereas the translation degrees of
freedom are described correctly (relativistically). Energetic levels,
connected with the internal degrees of freedom are very high. They are not
excited at the processes, which are considered in the quantum
electrodynamics and in the theory of atomic spectra. In these physical
processes one may ignore the internal degrees of freedom of the Dirac
particle. Thus, the correct description of the internal degrees of freedom
does not change anything in the quantum electrodynamics and in the theory of
atomic spectra. However, in the theory of the elementary particles, where
there are high energetic processes, we should take into account the internal
degrees of freedom of the Dirac particle. We are to describe them
relativistically.

We have produced the minimal modification of the classical Dirac particle $%
\mathcal{S}_{\mathrm{Dcl}}$, but the description of the Dirac particle $%
\mathcal{S}_{\mathrm{D}}$ remains to be not completely relativistic. This
defect should be eliminated also. We can make this, adding the eliminated
''quantum'' terms with transversal derivatives to the action for the
statistical ensemble $\mathcal{E}\left[ \mathcal{S}_{\mathrm{mDcl}}\right] $
of the modified classical Dirac particles $\mathcal{S}_{\mathrm{mDcl}}$.
This procedure is not unique. Besides, we cannot be sure, that the dynamic
equations in terms of the wave function appear to be linear.

Since 1925, when the Dirac equation has been invented, it was considered to
be the completely relativistic equation, describing the particles of spin
1/2 in the best way. Why was not the incompleteness of its relativity
discovered earlier? This question is very important for the correct
evaluation of the microcosm investigation strategy. The answer is rather
unexpected. The nonrelativistic features of the Dirac equation was
discovered in 1995 \cite{R95,R2001}, as soon as the Dirac particle became to
be investigated simply as a dynamic system (but not as the quantum dynamic
system). The quantum dynamic system distinguishes from the usual dynamic
system in its compatibility with the quantum principles. The quantum
principles restrict the way of description of the dynamic system. They
demand that the quantum dynamic system be described in terms of the world
function and dynamic equations be linear in these terms. The quantum
principles admit only linear transformation of \ dependent variables (wave
function).

The quantum principles impose the constraints on the description of a
quantum dynamic system and cumber its investigation. They do not admit to
eliminate $\gamma $-matrices, what is necessary for discovery of the dynamic
equation incompatibility with the principles of relativity. \bigskip \bigskip

\renewcommand{\theequation}{\Alph{section}.\arabic{equation}} %
\renewcommand{\thesection}{Appendix \Alph{section}} \setcounter{section}{0} %
\centerline{\Large \bf Mathematical Appendices}

\section{Solution of dynamic equations \newline
common to $\mathcal{S}_{\mathrm{Dcl}}$ and $\mathcal{S}_{\mathrm{mDcl}}$}

Dynamic equations generated by the action (\ref{f5.4}) have the form 
\begin{eqnarray}
\frac{\delta \mathcal{A}_{\mathrm{Dcl}}}{\delta x^{l}} &=&-\dot{p}%
_{l}=0,\qquad p_{l}=\text{const}  \label{A.1} \\
\frac{\delta \mathcal{A}_{\mathrm{Dcl}}}{\delta y^{l}} &=&\frac{\partial L}{%
\partial y^{l}}-\frac{d}{d\tau _{0}}\frac{\partial L}{\partial \dot{y}^{l}}%
-p_{l}=0,  \label{A.2} \\
\frac{\delta \mathcal{A}_{\mathrm{Dcl}}}{\delta \xi ^{l}} &=&\left( \frac{%
\partial L}{\partial \xi ^{s}}-\frac{d}{d\tau _{0}}\frac{\partial L}{%
\partial \dot{\xi}^{s}}\right) \left( \delta _{l}^{s}+\frac{\xi _{l}\xi ^{s}%
}{\xi _{k}\xi ^{k}}\right) =0  \label{A.3} \\
\frac{\delta \mathcal{A}_{\mathrm{Dcl}}}{\delta p_{l}} &=&\dot{x}^{l}-y^{l}=0
\label{A.4}
\end{eqnarray}
Let 
\[
y^{k}=\left\{ y^{0},\mathbf{y}\right\} ,\qquad p_{k}=\left\{ p_{0},\mathbf{p}%
\right\} =\left\{ p_{0},0\right\} ,\qquad \xi ^{k}=\left\{ 0,\mathbf{\xi }%
\right\} ,\qquad \mathbf{\xi }^{2}=1 
\]
Then for $l=1,2,3$ the dynamic equations (\ref{A.2}) can be written in the
form 
\begin{equation}
-\kappa _{0}m\mathbf{y}+\frac{\hbar Q}{2}(\mathbf{\xi }\times \mathbf{\dot{y}%
})-\frac{\hbar }{2}\frac{\partial Q}{\partial \mathbf{y}}(\mathbf{y}\times 
\mathbf{\dot{y}})\mathbf{\xi }+\frac{\hbar }{2}\frac{d}{d\tau _{0}}\left( Q(%
\mathbf{\xi }\times \mathbf{y})\right) =\mathbf{p}=0  \label{b6.1}
\end{equation}
where 
\begin{equation}
Q=Q\left( y\right) =\left( \sqrt{y^{s}y_{s}}(\sqrt{y^{s}y_{s}}+y^{0})\right)
^{-1}  \label{b6.2}
\end{equation}
For $\ l=0$ the dynamic equations (\ref{A.2}) can be written in the form 
\begin{equation}
-\kappa _{0}m\frac{y^{0}}{\sqrt{y^{s}y_{s}}}+\frac{\hbar }{2}\frac{\partial Q%
}{\partial y^{0}}(\mathbf{y}\times \mathbf{\dot{y}})\mathbf{\xi }=p_{0}
\label{b6.1a}
\end{equation}

The equation (\ref{A.3}) has the form 
\begin{equation}
-\hbar {\frac{\left( \dot{\mathbf{\xi }}\times \mathbf{z}\right) \times 
\mathbf{\xi }}{2(1+\mathbf{z\xi })}+}\hbar \left( {-}\frac{d}{d\tau _{0}}{{%
\frac{(\mathbf{\xi }\times \mathbf{z})}{2(1+\mathbf{z\xi })}}-\frac{(\dot{%
\mathbf{\xi }}\times \mathbf{\xi })\mathbf{z}}{2(1+\mathbf{z\xi })^{2}}%
\mathbf{z}}\right) \times \mathbf{\xi }+\hbar \frac{(\mathbf{y}\times 
\mathbf{\dot{y}})\times \mathbf{\xi }}{2}Q=0  \label{b6.5}
\end{equation}
After transformations this equation is reduced to the equation (see
Appendix\ B) 
\begin{equation}
\mathbf{\dot{\xi}}=-(\mathbf{y}\times \mathbf{\dot{y}})\times \mathbf{\xi }Q
\label{c7.1}
\end{equation}
which does not contain the vector $\mathbf{z}$. It means that $\mathbf{z}$
determines a fictitious direction in the space-time. Note that $\mathbf{z}$
in the action (\ref{c4.15}) for the system $\mathcal{S}_{\mathrm{D}}$ is
fictitious also, because there is only longitudinal derivative $%
j^{l}\partial _{l}$ in the term of the action (\ref{c4.15}) for $\mathcal{S}%
_{\mathrm{D}}$, which contains $\mathbf{z}$. This term is not changed at the
dynamical disquantization.

Constraint (\ref{f5.2c}) on the variables $\dot{x}^{k}$ remains to be valid
in the considered case. It takes the form 
\begin{equation}
\sqrt{y_{s}y^{s}}=1,\qquad y^{0}=\sqrt{1+\mathbf{y}^{2}}  \label{A.7}
\end{equation}
Taking into account the condition (\ref{A.7}) we obtain from (\ref{b6.2})
for quantities $Q$, $\partial Q/\partial y_{0}$, $\partial Q/\partial 
\mathbf{y}$ 
\begin{equation}
Q=\frac{1}{1+y_{0}},\qquad \frac{\partial Q}{\partial y_{0}}=-1,\qquad \frac{%
\partial Q}{\partial \mathbf{y}}=\frac{\mathbf{y}\left( 2+y_{0}\right) }{%
\left( 1+y_{0}\right) ^{2}}  \label{b6.11}
\end{equation}
The dynamic equations (\ref{c7.1}), (\ref{b6.1a}) and (\ref{b6.1}) can be
rewritten in the form 
\begin{equation}
\mathbf{\dot{\xi}}=-\frac{(\mathbf{y}\times \mathbf{\dot{y}})\times \mathbf{%
\xi }}{1+y_{0}}  \label{b6.7a}
\end{equation}

\begin{equation}
\kappa _{0}my_{0}+{\frac{\hbar }{2}}\left( \mathbf{y}\times \mathbf{\dot{y}}%
\right) \mathbf{\xi }=-p_{0}  \label{b6.7}
\end{equation}
\begin{equation}
-\kappa _{0}m\mathbf{y}+\frac{\hbar (\mathbf{\xi }\times \mathbf{\dot{y}})}{%
2\left( 1+y_{0}\right) }-\frac{\hbar }{2}\frac{\mathbf{y}\left(
2+y_{0}\right) }{\left( 1+y_{0}\right) ^{2}}(\mathbf{y}\times \mathbf{\dot{y}%
})\mathbf{\xi }+\frac{\hbar }{2}\frac{d}{d\tau _{0}}\left( \frac{(\mathbf{%
\xi }\times \mathbf{y})}{1+y_{0}}\right) =0  \label{e6.8}
\end{equation}

Equations (\ref{b6.7a}), (\ref{b6.7}), (\ref{e6.8}) admit the trivial
solution 
\begin{eqnarray}
\mathbf{y} &=&0,\qquad y^{0}=-\frac{\kappa _{0}p_{0}}{m}=\pm 1,\qquad 
\mathbf{\xi }=\mathbf{\xi }_{0}=\text{const,\qquad }\mathbf{\xi }^{2}=1
\label{e6.9} \\
x^{k} &=&\left\{ -\varepsilon \kappa _{0}\tau _{0},X^{1},X^{2},X^{3}\right\}
,\qquad X^{\alpha }=\text{const},\;\;\alpha =1,2,3,\qquad \varepsilon =\text{%
sgn}p_{0}  \label{e6.10}
\end{eqnarray}
In this case the internal degrees of freedom are not excited, and the world
line of the classical Dirac particle $\mathcal{S}_{\mathrm{Dcl}}$ is a
timelike straight in the space-time.

For solution of dynamic equations it is important to take into account that
three 3-vectors $\mathbf{\xi }$,$\mathbf{y},\mathbf{\dot{y}}$ are orthogonal
between themselves 
\begin{equation}
\mathbf{\xi y}=0,\qquad \mathbf{\xi \dot{y}}=0,\qquad \mathbf{y\dot{y}}%
=0,\qquad \mathbf{y}^{2}=\gamma ^{2}-1=\text{const},  \label{A.8}
\end{equation}
To prove relations (\ref{A.8}) we transform the relation (\ref{e6.8}),
calculating the last term. 
\begin{equation}
-\kappa _{0}m\mathbf{y}+\frac{\hbar (\mathbf{\xi }\times \mathbf{\dot{y}})}{%
\left( 1+y_{0}\right) }-\frac{\hbar }{2}\frac{\mathbf{y}\left(
2+y_{0}\right) }{\left( 1+y_{0}\right) ^{2}}(\mathbf{y}\times \mathbf{\dot{y}%
})\mathbf{\xi }+\frac{\hbar }{2}\frac{(\mathbf{\dot{\xi}}\times \mathbf{y})}{%
1+y_{0}}=0  \label{A.9}
\end{equation}

Using (\ref{b6.7a}), we eliminate $\mathbf{\dot{\xi}}$ and obtain 
\begin{equation}
-\kappa _{0}m\mathbf{y}+\frac{\hbar (\mathbf{\xi }\times \mathbf{\dot{y}})}{%
\left( 1+y_{0}\right) }-\frac{\hbar }{2}\frac{\mathbf{y}\left(
2+y_{0}\right) }{\left( 1+y_{0}\right) ^{2}}(\mathbf{y}\times \mathbf{\dot{y}%
})\mathbf{\xi }-\frac{\hbar }{2}\frac{\left( (\mathbf{y}\times \mathbf{\dot{y%
}})\times \mathbf{\xi }\right) \times \mathbf{y}}{\left( 1+y_{0}\right) ^{2}}%
=0  \label{A.10}
\end{equation}
Calculating the double vector product in the last term of (\ref{A.10}) and
combining it with the second term, we obtain 
\begin{equation}
\frac{\hbar }{\left( 1+y_{0}\right) }\left( \mathbf{\xi }+\frac{1}{2}\frac{%
\left( \mathbf{\xi y}\right) \mathbf{y}}{\left( 1+y_{0}\right) }\right)
\times \mathbf{\dot{y}}-\left( \frac{\hbar }{2}\frac{\left( 2+y_{0}\right) }{%
\left( 1+y_{0}\right) ^{2}}(\mathbf{y}\times \mathbf{\dot{y}})\mathbf{\xi }%
+\kappa _{0}m\right) \mathbf{y}=0  \label{A.11}
\end{equation}
It follows from (\ref{A.11}) that the vector $\mathbf{y}$ is orthogonal to
vectors $\mathbf{\dot{y}}$ and $\mathbf{\xi }+\frac{1}{2}\left(
1+y_{0}\right) ^{-1}\left( \mathbf{\xi y}\right) \mathbf{y}$, provided 
\begin{equation}
\frac{\hbar }{2}\frac{\left( 2+y_{0}\right) }{\left( 1+y_{0}\right) ^{2}}(%
\mathbf{y}\times \mathbf{\dot{y}})\mathbf{\xi }+\kappa _{0}m\neq 0
\label{A.12}
\end{equation}

It means that the last two relations (\ref{A.8}) are fulfilled. Besides 
\begin{equation}
\left( 1+\frac{\mathbf{y}^{2}}{2\left( 1+y_{0}\right) }\right) \left( 
\mathbf{\xi y}\right) =0  \label{A.14}
\end{equation}
and the first relation (\ref{A.8}) takes place, as far as according to the
last relation (\ref{A.8}) $\gamma ^{2}>1$, and 
\begin{equation}
1+\frac{\mathbf{y}^{2}}{2\left( 1+y_{0}\right) }=\frac{\gamma +1}{2}\neq 0
\label{A.15}
\end{equation}
(In the case $\gamma ^{2}=1$ we have the trivial solution $\mathbf{y}=0$).
Differentiating of the first relation (\ref{A.8}) and using the relation (%
\ref{b6.7a}) for elimination of $\mathbf{\dot{\xi}}$, we obtain 
\begin{equation}
\left( \mathbf{\xi \dot{y}}\right) =-\left( \mathbf{\dot{\xi}y}\right) =%
\mathbf{y}\frac{(\mathbf{y}\times \mathbf{\dot{y}})\times \mathbf{\xi }}{%
1+y_{0}}=-\frac{\mathbf{y}^{2}\left( \mathbf{\xi \dot{y}}\right) }{1+y_{0}}%
=-\left( \mathbf{\xi \dot{y}}\right) \left( \gamma -1\right)   \label{A.16}
\end{equation}
As far as $\gamma ^{2}\geq 1$, the second relation (\ref{A.8}) follows from (%
\ref{A.16}). Thus, if the condition (\ref{A.12}) is fulfilled the relations (%
\ref{A.8}) are also fulfilled.

Let now the condition (\ref{A.12}) be violated. Eliminating $(\mathbf{y}%
\times \mathbf{\dot{y}})\mathbf{\xi }$ from the relation 
\begin{equation}
\frac{\hbar }{2}\frac{\left( 2+y_{0}\right) }{\left( 1+y_{0}\right) ^{2}}(%
\mathbf{y}\times \mathbf{\dot{y}})\mathbf{\xi }+\kappa _{0}m=0  \label{A.17}
\end{equation}
by means of the dynamic equation (\ref{b6.7}), we obtain 
\begin{equation}
\left( 2+y_{0}\right) \left( -p_{0}-\kappa _{0}my_{0}\right) +\kappa
_{0}m\left( 1+y_{0}\right) ^{2}=0  \label{A.18}
\end{equation}
Resolving (\ref{A.18}) with respect to $y_{0}$, we obtain 
\begin{equation}
y_{0}=\frac{\kappa _{0}m}{p_{0}}-2=\text{const}  \label{A.19}
\end{equation}

In this case it follows from (\ref{A.11}) 
\begin{equation}
\mathbf{\dot{y}}=\alpha \left( \mathbf{\xi }+\frac{1}{2}\frac{\left( \mathbf{%
\xi y}\right) \mathbf{y}}{\left( 1+y_{0}\right) }\right)  \label{A.20}
\end{equation}
where $\alpha $ is some real number. Besides, it follows from (\ref{A.7})
that $\mathbf{y}^{2}=$const and hence the last two relations (\ref{A.8}) are
fulfilled. Multiplying (\ref{A.20}) by $\mathbf{y}$\textbf{, }we obtain 
\begin{equation}
\mathbf{\dot{y}y}=\alpha \left( 1+\frac{\mathbf{y}^{2}}{2\left(
1+y_{0}\right) }\right) \left( \mathbf{\xi y}\right) =0  \label{A.21}
\end{equation}
Hence, either $\alpha =0$, or $\left( \mathbf{\xi y}\right) =0$. If $\alpha
=0$, then it follows from (\ref{A.20}), that $\mathbf{\dot{y}}=0$. But the
condition $\mathbf{\dot{y}}=0$ is compatible with the relation (\ref{A.17})
only if $m=0$. Hence, $\alpha \neq 0$, and it follows from (\ref{A.21}),
that $\left( \mathbf{\xi y}\right) =0$, and we obtain from (\ref{A.20}) 
\begin{equation}
\mathbf{\dot{y}}=\alpha \mathbf{\xi }  \label{A.22}
\end{equation}
In the case (\ref{A.22}) $(\mathbf{y}\times \mathbf{\dot{y}})\mathbf{\xi }=0$%
, and the relation (\ref{A.22}) is compatible with the relation (\ref{A.17})
only if $m=0$.

We suppose that $m\neq 0$, and the relation (\ref{A.17}) is fulfilled never.
The opposite relation (\ref{A.12}) is fulfilled always, as well as the
relations (\ref{A.8}).

According to constraints (\ref{A.8}) the dynamic equation (\ref{b6.7a})
turns into 
\begin{equation}
\mathbf{\dot{\xi}}=0,\qquad \mathbf{\xi }=\text{const,}\qquad \mathbf{\xi }%
^{2}=1  \label{A.23}
\end{equation}
We choose the coordinate system in such a way that 
\begin{equation}
\mathbf{\xi }=\left\{ 0,0,\varepsilon _{0}\right\} ,\qquad \varepsilon
_{0}=\pm 1  \label{A.24}
\end{equation}
According to (\ref{A.8}) vectors $\mathbf{y}$ and $\mathbf{\dot{y}}$ in this
coordinate system can be written as follows 
\begin{eqnarray}
\mathbf{y} &=&\left( y^{1},y^{2},0\right) =\left\{ \sqrt{\gamma ^{2}-1}\cos
\Phi ,\sqrt{\gamma ^{2}-1}\sin \Phi ,0\right\} ,\qquad \gamma =\text{const},
\label{e6.21} \\
\mathbf{\dot{y}} &=&\left( \dot{y}^{1},\dot{y}^{2},0\right) =\left\{ -\sqrt{%
\gamma ^{2}-1}\omega \sin \Phi ,\sqrt{\gamma ^{2}-1}\omega \cos \Phi
,0\right\} ,\qquad \omega =\frac{d\Phi }{d\tau _{0}}  \label{e6.22}
\end{eqnarray}
It follows from the last relation (\ref{A.8}) and (\ref{A.7}), that 
\begin{equation}
y^{0}=\gamma =\text{const }  \label{A.25a}
\end{equation}
Substituting (\ref{A.24}), (\ref{e6.21}), and (\ref{e6.22}) in the dynamic
equation (\ref{b6.7}), we obtain 
\begin{equation}
\left( \mathbf{y}\times \mathbf{\dot{y}}\right) \mathbf{\xi }=\varepsilon
_{0}\left( \gamma ^{2}-1\right) \omega \mathbf{\qquad }\gamma =-\frac{\kappa
_{0}p_{0}}{m}-\varepsilon _{0}\kappa _{0}\frac{\hbar }{2m}\left( \gamma
^{2}-1\right) \omega  \label{A.26}
\end{equation}
Then the second relation (\ref{A.26}) leads to the conclusion that $\omega =$%
const.

Substituting (\ref{A.24}), and (\ref{A.26}) in the dynamic equation (\ref
{A.11}), we obtain 
\begin{equation}
\frac{\hbar }{\left( 1+y_{0}\right) }\mathbf{\xi }\times \mathbf{\dot{y}}%
-\left( \frac{\hbar }{2}\frac{\left( 2+y_{0}\right) }{\left( 1+y_{0}\right)
^{2}}\varepsilon _{0}\left( \gamma ^{2}-1\right) \omega +\kappa _{0}m\right) 
\mathbf{y}=0  \label{A.27}
\end{equation}
It follows from (\ref{e6.21}), (\ref{e6.22}) and (\ref{A.24}), that 
\begin{equation}
\mathbf{\xi }\times \mathbf{\dot{y}}=-\varepsilon _{0}\omega \mathbf{y}
\label{A.28}
\end{equation}
Substituting (\ref{A.24}) and (\ref{A.25a}) in (\ref{A.27}), we obtain 
\begin{equation}
-\varepsilon _{0}\omega \mathbf{y}-\left( \frac{\left( 2+\gamma \right) }{%
2\left( 1+\gamma \right) }\varepsilon _{0}\left( \gamma ^{2}-1\right) \omega
+\frac{\kappa _{0}m}{\hbar }\left( 1+\gamma \right) \right) \mathbf{y}=0
\label{A.29}
\end{equation}
Resolving (\ref{A.29}) with respect to $\omega $, we obtain 
\begin{equation}
\omega =-2\kappa _{0}\varepsilon _{0}\frac{m}{\gamma \hbar }  \label{A.30}
\end{equation}

It follows from the second equation (\ref{A.26}) and (\ref{A.30}), that 
\begin{equation}
p_{0}=-\frac{\kappa _{0}m}{\gamma }  \label{A.31}
\end{equation}
The solution (\ref{e6.21}) has the form 
\begin{equation}
y^{k}=\left\{ \gamma ,\sqrt{\gamma ^{2}-1}\cos \left( \omega \tau _{0}+\phi
\right) ,\sqrt{\gamma ^{2}-1}\sin \left( \omega \tau _{0}+\phi \right)
,0\right\} ,  \label{A.32}
\end{equation}
where $\omega $ is determined by the relation (\ref{A.30}), and $\gamma
^{2}\geq 1$, $\phi $, are constants.

If $\gamma =1$ we obtain 
\begin{equation}
y^{k}=\left\{ 1,0,0,0\right\} ,  \label{A.33}
\end{equation}
This solution coincides with the trivial solution (\ref{e6.9}), (\ref{e6.10}%
).

\section{Transformation of equation for \newline
variable $\mathbf{\protect\xi }$}

Multiplying equation (\ref{b6.5}) by $(1+\mathbf{z\xi })$ and keeping in
mind $\mathbf{\xi }^{2}=1$ and $\mathbf{z}^{2}=1$, we obtain 
\begin{equation}
\mathbf{\xi }\times \left( -\dot{\mathbf{\xi }}\times \mathbf{z}+{\frac{(%
\mathbf{z}\dot{\mathbf{\xi }})}{2(1+\mathbf{z\ \xi })}}\mathbf{\xi }\times 
\mathbf{z-}\frac{\mathbf{\dot{\xi}}(\mathbf{\xi }\times \mathbf{z})}{2(1+%
\mathbf{z\ \xi })}\mathbf{z}-{\frac{(1+\mathbf{z\ \xi })}{2}}\mathbf{b}%
\right) =0,\qquad \mathbf{b}=-(\mathbf{y}\times \mathbf{\dot{y}})Q
\label{f9.1}
\end{equation}
Two middle terms could be represented as the double vector product 
\begin{equation}
\mathbf{\xi }\times \left( -\dot{\mathbf{\xi }}\times \mathbf{z}+{\frac{1}{%
2(1+\mathbf{z\ \xi })}}\left( \mathbf{\dot{\xi}}\times \left( {(\mathbf{\xi }%
\times \mathbf{z})\times }\mathbf{z}\right) \right) -{\frac{(1+\mathbf{z\
\xi })}{2}}\mathbf{b}\right) =0  \label{f9.2}
\end{equation}
This equation can be rewritten in the form 
\begin{equation}
\mathbf{\xi }\times \left( \dot{\mathbf{\xi }}\times \left( -\mathbf{z+}{%
\frac{\left( \mathbf{z\xi }\right) \mathbf{z-\xi }}{2(1+\mathbf{z\ \xi })}}%
\right) -{\frac{(1+\mathbf{z\ \xi })}{2}}\mathbf{b}\right) =0  \label{f9.3}
\end{equation}
Now calculating the double vector products and taking into account that $%
\mathbf{\xi \dot{\xi}}=0$, we obtain 
\[
\left( \mathbf{\dot{\xi}}\left( -\mathbf{z\xi +}{\frac{\left( \mathbf{z\xi }%
\right) ^{2}\mathbf{-}1}{2(1+\mathbf{z\ \xi })}}\right) \right) -{\frac{(1+%
\mathbf{z\ \xi })}{2}}\left( \mathbf{\xi }\times \mathbf{b}\right) =0 
\]
\[
\left( \mathbf{\dot{\xi}}\left( -\mathbf{z\xi +}{\frac{\left( \mathbf{z\xi }%
\right) \mathbf{-}1}{2}}\right) \right) -{\frac{(1+\mathbf{z\ \xi })}{2}}%
\left( \mathbf{\xi }\times \mathbf{b}\right) =0 
\]
\begin{equation}
-\dot{\mathbf{\xi }}-\left( \mathbf{\xi }\times \mathbf{b}\right) =0
\label{f9.5}
\end{equation}
or using designation (\ref{f9.1}) 
\begin{equation}
\dot{\mathbf{\xi }}=\left( \mathbf{\xi }\times (\mathbf{y}\times \mathbf{%
\dot{y}})\right) Q  \label{f9.6}
\end{equation}

\end{document}